\documentclass[prb,aps,twocolumn,floatfix]{revtex4}
\usepackage{graphics}
\begin{document}
\title{BCS-BEC crossover of collective excitations in two-band superfluids}
\author{M. Iskin and C. A. R. S{\'a} de Melo}
\affiliation{School of Physics, Georgia Institute of Technology, Atlanta, Georgia 30332, USA}
\date{\today}

\begin{abstract}
We use the functional integral approach to study low energy 
collective excitations in a continuum model of neutral 
two-band superfluids at $T=0$ for all couplings with a separable pairing interaction.
In the long wavelength and low frequency limit, we recover 
Leggett's analytical results in weak coupling 
(BCS) for $s$-wave pairing, and further obtain analytical results in  
strong coupling (BEC) for both two and three dimensional systems. 
We also analyse numerically the behavior of
the out-of-phase {\it exciton} (finite frequency) mode and 
the in-phase {\it phonon} (Goldstone)
mode from weak to strong coupling limits, including the crossover region. 
In principle, the evolution of Goldstone and finite 
frequency modes from weak to strong coupling may be
accessible experimentally in the superfluid phase of neutral Fermi atomic gases, 
and could serve as a test of the validity of the theoretical 
analysis and approximations proposed here.

\pacs{PACS:}PACS: 74.40.+k,74.20.Fg,05.30.Fk,03.75.Kk
\end{abstract}
\maketitle

\section{Introduction}
A two-band model for superfluidity was introduced by Suhl et al.~\cite{suhl} in 1959 soon 
after the BCS theory in order to allow for the possibility of multiple
band crossing through the Fermi surface. Suhl et al.~\cite{suhl} observed 
that a larger number of energy bands crossing the Fermi surface could
increase the overall electron state density and lead to the onset of 
additional interactions. Since then, this model has been used to describe 
high temperature superconductivity 
in copper oxides and more recently it has been used in connection to $\rm {MgB_2}$~\cite{amy,kortus}. 

In the seminal work by Leggett~\cite{leggett}, the existence of collective 
phase modes in two-band superfluids has been predicted in both neutral and charged systems. 
Within the weak coupling (BCS) limit, Leggett showed that if 
$s$-wave interactions are attractive in both bands, 
an undamped long-wavelength {\it exciton} (finite frequency) mode, 
as well as an undamped long-wavelength {\it phonon} (Goldstone) mode may exist.
These modes were further studied theoretically~\cite{manasori2,sharapov} in the BCS limit, however, 
there was no experimental evidence of their existence until manifestations of 
two-gap behaviour~\cite{mgb21,mgb22,mgb23,mgb24,ponomanev} were found in ${\rm MgB_2}$. 

However, it has not been possible to study the evolution of collective spectra 
from weak (BCS) to strong coupling (BEC) limit until very recently.
Advances in experiments with neutral Fermi gases enabled the
tunning and control of two-particle interactions between atoms in different 
hyperfine states by using Feshbach resonances~\cite{regall,ohara2}.
This kind of control is not fully present in standard fermionic 
condensed matter systems, and has hindered the development 
of experiments that could probe systematically
the effects of strong correlations as a function of coupling or density of fermions.
It was thought theoretically for many years that a weakly 
coupled (BCS) superfluid could evolve smoothly  
into the limit of tightly bound pairs which undergo Bose-Einstein 
condensation (BEC)~\cite{eagles,leggett2, nsr,carlos,perali,ohashi,holland}. 
It was not until recently that the first experimental evidence that hyperfine states of $^{40} {\rm K}$ 
(Ref.~\cite{K1,K2}) and $^6{\rm Li}$ (Ref.~\cite{litium1,litium3,litium4,litium5}) 
can form weakly and tightly bound atom pairs (Cooper pairs), 
when the magnetic field is swept through an $s$-wave Feshbach resonance.

Considering these recent findings in condensed matter systems and advances 
in atomic physics experiments,
we expand Leggett's calculation of collective modes in neutral two-band 
($s$-wave) superfluids to all couplings 
by following a similar one-band approach \cite{jan}. 
These collective modes for two-band $s$-wave 
systems are undamped in the low-frequency and low-momentum limits, provided that 
the two-quasiparticle threshold is not reached.
We present results of the evolution of the finite frequency and
Goldstone's modes from weak coupling (BCS) and to strong coupling (BEC) limits, 
and discuss briefly the possibility of observing these modes 
in experiments involving multi-component ultracold atomic Fermi gases. 
While it is still a matter of debate that extentions of
Eagles'~\cite{eagles}, Leggett's~\cite{leggett2}, 
and Nozieres and Schmitt-Rink's~\cite{nsr} (NSR) suggestions are good quantitative 
description of the crossover phenomena, the evolution of collective modes 
can serve as a test of these ideas.

The rest of the paper is organized as follows. 
In section~\ref{sec:effective-action-method}, 
we discuss the effective action and the saddle point approximation for 
a two-band continuum Hamiltonian with attractive interactions in the $s$-wave 
channel and a Josephson interband coupling term. We analyse the effects of 
Gaussian fluctuations in section~\ref{sec:gaussian-fluctuations}, 
and derive effective amplitude and phase actions, from which the in-phase 
(Goldstone) and out-of-phase (finite frequency) collective mode spectra 
are calculated in section~\ref{sec:collective-modes}. The evolution of 
these modes is analysed as a function of 
interaction strengths from the BCS to the BEC limit both analytically and numerical 
in section~\ref{sec:analytical-results}. 
In section~\ref{sec:conclusions}, we summarize our conclusions and
propose that ultra-cold Fermi atoms can be used to test our results 
regarding the evolution of the finite frequency and
Goldstone modes in two-band superfluids. Finally, in Appendix A, 
we present details of the matrix elements involved in the 
phase and amplitude effective actions, while in Appendix B, 
we show the long wavelength expansion coefficients needed to 
evaluate the phase modes.

\section{Effective Action Method}
\label{sec:effective-action-method}

\begin{figure} [ht]
\centerline{\scalebox{0.55}{\includegraphics{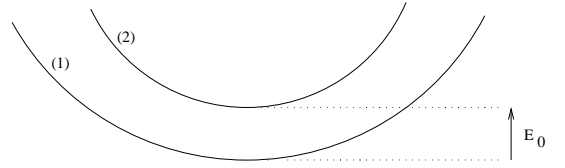}}}
\caption{\label{bands} Schematic figure of two-bands with reference energies 
$\varepsilon_{1,0}=0$ for the band 1 and
$\varepsilon_{2,0} = E_0$ for band 2.}
\end{figure}

A Hamiltonian for multi-band (or multi-component) superfluids with singlet pairing can be written as
\begin{eqnarray}
H&=&\sum_{n,\mathbf{k},\sigma}\xi_n(\mathbf{k})a_{n,\mathbf{k},\sigma}^\dagger a_{n,\mathbf{k},\sigma} 
+ \sum_{n,m,\mathbf{q}} U_{nm}(\mathbf{q})\rho_{n, \mathbf{q}} \rho_{m,-\mathbf{q}}	\nonumber \\ 
&+& \sum_{n,m,\mathbf{k},\mathbf{k'},\mathbf{q}}
V_{nm}(\mathbf{k},\mathbf{k'}) b_{n,\mathbf{k},\mathbf{q}}^\dagger b_{m,\mathbf{k'},\mathbf{q'}}	
\label{hamiltonian}
\end{eqnarray}
where the indices $n$ and $m$ label different bands (or components), 
$\sigma$ labels the spins (or pseudo-spins), and $\mathbf{k}$ labels the momentum.
In addition, $a_{n,\mathbf{k},\uparrow}^\dagger$ is the fermion creation operator,
$
\rho_{n,\mathbf{q}}=\sum_{\mathbf{k}, \sigma} a_{n,\mathbf{k}-\mathbf{q},\sigma}^\dagger a_{n,\mathbf{k},\sigma}
$ is the density operator, and 
$b_{n,\mathbf{k},\mathbf{q}}^\dagger = 
a_{n,\mathbf{k}+\mathbf{q}/2,\uparrow}^\dagger a_{n,-\mathbf{k}+\mathbf{q}/2,\downarrow}^\dagger$ 
is the pair creation operator. 
Here, $\xi_n(\mathbf{k})=\varepsilon_n(\mathbf{k})-\mu$, where 
\begin{equation}
\varepsilon_n(\mathbf{k})=\varepsilon_{n,0}+\mathbf{k}^2/2m_n
\end{equation}
(with $\hbar = 1$) is the kinetic energy of fermions. 
The reference energies are $\varepsilon_{1,0}=0$ and 
$\varepsilon_{2,0} = E_0 > 0$ (see Fig.~\ref{bands});  
$m_n$ is the effective mass for the $n^{th}$ band.
Furthermore, the terms $U_{nm}$ and $V_{nm}$ correspond to the Coulomb and 
pairing interaction matrix elements, respectively.
Since in this paper we are interested only in the neutral case, 
we will ignore Coulomb interactions, and
consider only the pairing term. This choice is more appropriate to 
ultracold atomic Fermi gases, while the inclusion of Coulomb terms is more appropriate 
in the case of superconductors. The discussion of the charged 
case is postponed to a future manuscript. For the neutral case, we take 
\begin{equation}
V_{nm}(\mathbf{k},\mathbf{k'})=V_{nm}\Gamma_n^*(\mathbf{k})\Gamma_m(\mathbf{k'})
\end{equation}
to be separable, and we consider in general a two-band system 
with distinct intraband interactions $V_{11}$ and $V_{22}$, 
and interband interactions $V_{12}$ and $V_{21}$. Notice that 
the off-diagonal terms $V_{12}$ and $V_{21}$ play 
the role of Josephson coupling terms. For the purpose of this 
paper we will consider all $V_{nm}$ to be negative. The $\Gamma_n(\mathbf{k})$
coefficients are symmetry factors characterizing the chosen angular momentum channel.

In the imaginary-time functional integration formalism 
($\beta=1/T$, $\hbar=k_B=1$), the partition function is 
written as
\begin{equation}
Z=\int D[a^\dagger,a]e^{-S}
\end{equation}
with an action given by
\begin{equation}
S = \int_0^\beta d\tau
\left[ 
\sum_{n,\mathbf{k},\sigma}a_{n,\mathbf{k},\sigma}^\dagger(\tau)(\partial_\tau)a_{n,\mathbf{k},\sigma}(\tau) 
+ H(\tau) 
\right].
\end{equation}
The Hamiltonian from Eq.~(\ref{hamiltonian}) can be re-written in the form
\begin{eqnarray}
H(\tau)&=& \sum_{n,m,\mathbf{k},\sigma}\xi_n(\mathbf{k})a_{n,\mathbf{k},\sigma}^\dagger(\tau)a_{n,\mathbf{k},\sigma}(\tau) \nonumber \\
&+& \sum_{n,m,\mathbf{q}}\mathbf{B}_n^\dagger(\mathbf{q},\tau)V_{nm}\mathbf{B}_m(\mathbf{q},\tau),
\end{eqnarray} 
with $\mathbf{B}_n(\mathbf{q},\tau)=\sum_{\mathbf{k}}\Gamma_n(\mathbf{k})b_{n,\mathbf{k},\mathbf{q}}$.
We first introduce the Nambu spinor 
$
\psi^\dagger_n(p)=( \begin{array}{cc}a_{n,p\uparrow}^\dagger, a_{n,-p\downarrow} \end{array}),
$
where $p=(\mathbf{k},iw_\ell)$ is used to denote both 
momentum and Matsubara frequencies ($w_\ell=(2\ell+1)\pi/\beta$).
Furthermore, we use the Hubbard-Stratanovich transformation 
\begin{widetext}
\begin{eqnarray}
\exp\big\lbrace-\sum_{n,m,q}\mathbf{B}_n^\dagger(q)V_{nm}\mathbf{B}_m(q)\big\rbrace
= \int D[\Phi^\dagger,\Phi]\exp\big\lbrace\sum_{n,m,q}\Phi_n^\dagger(q)g_{nm}\Phi_m(q) + 
\sum_{n,q}\left[\mathbf{B}_n^\dagger(q)\Phi_n(q) + h.c. \right]\big\rbrace
\end{eqnarray}
\end{widetext}
to decouple the fermionic degrees of freedom at the expense 
of introducing the bosonic fields $\Phi_n (q)$, with $q=(\mathbf{q},iv_\ell)$ where $v_\ell= 2\ell\pi/\beta$. 
The tensor $g_{mn}$ associated with the bosonic pairing fields $\Phi_m(q)$ can be written as
\begin{equation}
\left( \begin{array}{cc} g_{11} & g_{12} \\ g_{21} & g_{22} \end{array}\right) =
\frac{1}{\det \mathbf{V}}\left( \begin{array}{cc} V_{22} & -V_{12} \\ -V_{21} & V_{11} 
\end{array}\right), 
\label{inverseV}
\end{equation}
where $\det \mathbf{V}=V_{11}V_{22}-V_{12}V_{21}$.
Performing an integration over the fermionic part 
($D[\psi^\dagger,\psi]$) leads to
\begin{eqnarray}
S&=&-\beta\sum_{n,m,q} \Phi_n^\dagger(q)g_{nm}\Phi_m(q) \nonumber \\
&+& \sum_{n,p,p'}\left[ \xi_n(\mathbf{k})\delta_{p,p'}-\rm {Tr}\ln\mathbf{G}_n^{-1}\right]. \label{}
\end{eqnarray}
Here, the inverse Nambu matrix is 
\begin{eqnarray}
\mathbf{G}_n^{-1} &=& \Phi_n(-q)\Gamma_n^*(\frac{p+p'}{2})\sigma_-  
+ \Phi_n^\dagger(q)\Gamma_n(\frac{p+p'}{2})\sigma_+ \nonumber \\
&+& \left[iw_\ell \sigma_0-\xi_n(\mathbf{k})\sigma_3 \right] \delta_{p,p'}, 
\end{eqnarray}
where $\sigma_{\pm}=(\sigma_1\pm\sigma_2)/2$ and $\sigma_{i}$ are the Pauli spin matrices.

We choose an approximation scheme where the field $\Phi_n(q)$ is written as a 
sum of a $\tau$-independent (stationary) part and 
a $\tau$-dependent contribution 
\begin{equation}
\Phi_n(p-p')= \Phi_n(q)=\Delta_{n,0}\delta_{q,0}+\Lambda_n(q).
\end{equation}
We first write the inverse Nambu matrix in terms of two matrices:
\begin{equation}
\mathbf{G}_n^{-1} = \mathbf{G_{n,0}}^{-1}(\mathbf{1}+\mathbf{G_{n,0}}\mathbf{G_{n,1}}^{-1}).
\end{equation}
The first one is the saddle point inverse Nambu matrix given by
\begin{equation} 
\mathbf{G_{n,0}}^{-1} = 
\delta_{p,p'}\left( \begin{array}{cc} iw-\xi_n(\mathbf{k}) & \Delta_{n,0}^*\Gamma_n(\mathbf{p}) \\ 
\Delta_{n,0}\Gamma_n^*(\mathbf{p}) & iw+\xi_n(\mathbf{k}) \end{array}\right),
\end{equation}
and the second one is the fluctuation matrix 
\begin{equation}
\mathbf{G_{n,1}}^{-1} = 
\left( \begin{array}{cc} 0 & \Lambda_n^*(\mathbf{-q})\Gamma_n(\frac{\mathbf{p}+\mathbf{p'}}{2}) \\ 
\Lambda_n(\mathbf{q})\Gamma_n^*(\frac{\mathbf{p}+\mathbf{p'}}{2}) & 0 \end{array}\right),
\end{equation}
used in the expansion of the natural logarithm of
\begin{equation}
\rm{Tr}\ln\mathbf{G}_n^{-1} = \rm{Tr}\ln \mathbf{G_{n,0}} 
- \sum_{i=1}^\infty \frac{(-1)^i}{i}\rm{Tr}(\mathbf{G_{n,0}}\mathbf{G_{n,1}}^{-1})^i. \nonumber
\end{equation}

Within this approximation, we expand the action to second order in the fluctuation field $\Lambda_n(q)$.
This procedure produces an effective action of the form 
$S = S_0 + S_2$, where  
\begin{eqnarray*}
S_0 &=& -\beta \sum_{n,m}g_{n,m}\Delta^*_{n,0}\Delta_{m,0} \nonumber \\
&+& \sum_{n,p,p'}\left[\xi_n(\mathbf{k})\delta_{p,p'}
-\rm{Tr}\ln\mathbf{G}_{n,0}^{-1}\right]
\end{eqnarray*}
is the effective saddle point action, and 
$$
S_2 = \sum_{n,p_i} \mathbf{G_{n,0}}(12) 
\mathbf{G_{n,1}}^{-1}(23) 
\mathbf{G_{n,0}}(34) \mathbf{G_{n,1}}^{-1}(41)
$$
is the Gaussian correction to it. 
The notation $(ij)$ in the $\mathbf{G}$ matrices is
understood as the momentum labels $(p_i, p_j)$.
An important comment about $S_0$ is in order. 
Writing $\Delta_{n,0}$ in terms of its amplitude and phase
\begin{equation}
\Delta_{n,0} = \vert \Delta_{n,0} \vert \exp (i \varphi_n ),
\end{equation}
the first term of $S_0$ becomes the sum of two contributions:
the standard band-diagonal terms
$-\beta \sum_n g_{nn} \vert \Delta_{n,0} \vert^2$, 
and the band-off-diagonal
$-2 \beta g_{12} \vert \Delta_{1,0} \vert \vert \Delta_{2,0} \vert
\cos (\varphi_2 - \varphi_1)$ 
corresponding to the Josephson coupling between bands.
Since $g_{12} = - V_{12}/\det \mathbf{V}$, with $\det \mathbf{V} >  0$ 
and all $V_{nm} <  0$, 
the saddle point thermodynamic potential
$\Omega_0 = S_0/\beta$, has its quadratic term of the form
$
(
\vert V_{22} \vert \vert \Delta_{1,0} \vert^2
+\vert V_{11} \vert \vert \Delta_{2,0} \vert^2
-2 \vert V_{12} \vert \vert \Delta_{1,0} \vert \vert \Delta_{2,0} \vert
\cos (\varphi_2 - \varphi_1) 
)/ \det \mathbf{V},
$
which shows explicity the Josephson energy. For the case chosen,
where all $V_{nm}$ are negative, the in-phase $\varphi_2 = \varphi_1$
is the only stable solution. However, if $V_{12}$ were positive, another
stable solution would appear where $\varphi_2 = \varphi_1 + \pi$. 
This is the so-called $\pi$-phase solution, which we 
will not discuss here.

From the stationary condition 
$\partial S_{0}/\partial \Delta^*_n(q) = 0$ we obtain the  
order parameter equations
\begin{eqnarray}
\Delta_{m,0}=
-\sum_{n,\mathbf{k'}}V_{nm}\Gamma_n(\mathbf{k'})
\Delta_{n,0}\Gamma_n(\mathbf{k'})
\frac{\tanh \frac{\beta E_n(\mathbf{k'})}{2}}{2E_n(\mathbf{k'})} 
\label{orderparameter}
\end{eqnarray}
where $E_n(\mathbf{k})=(\xi_n^2(\mathbf{k})+
\vert\Delta_n(\mathbf{k})\vert^2)^{\frac{1}{2}}$ is the quasiparticle energy spectrum.
Note that the order parameter 
\begin{equation}
\Delta_n(\mathbf{k}) = \Delta_{n,0}\Gamma_n(\mathbf{k})
\end{equation}
is a separable function of temperature $T$ and momentum $\mathbf{k}$. 

In this manuscript, we focus on $s$-wave superfluids, and thus 
consider only the zero angular momentum channel of
the interaction $V_{nm} (\mathbf{k}, \mathbf{k'})$.
In addition, instead of taking $\Gamma_n(\mathbf{k})= 1$ (independent of ${\bf k}$)
which would cause ultraviolet divergences (logarithmic in two dimensions and linear in 
three dimensions) in the integrations over momentum for the order parameter equation, 
we take 
\begin{equation}
\Gamma_n(\mathbf{k})=1/(1+k/k_{n,0})^{1/2}
\end{equation}
as the corresponding symmetry factor in two dimensions (2D), and 
\begin{equation}
\Gamma_n(\mathbf{k})=1/(1+k/k_{n,0})
\end{equation}
as the corresponding symmetry factor in three dimensions (3D). 
Here $k_{n,0}\sim R_{n,0}^{-1}$, where $R_{n,0}$ plays 
the role of the interaction range in real space, sets the scale at large momenta, and
it is necessary to produce the physically correct behaviour of a generic
interaction at short wavelengths. In the case of 2D, 
$V_{nn} (\mathbf{k}, \mathbf{k'}) \sim 1$ for small $\mathbf{k}$ and 
$\mathbf{k'})$, while 
$V_{nn} (\mathbf{k}, \mathbf{k'}) \sim 1/\sqrt{k k'}$ for large $\mathbf{k}$ and 
$\mathbf{k'})$, when a generic real potential is used~\cite{richard}. 
There is no ultraviolet divergence in our theory since the momentum integrations
always produce finite results. This choice of interactions has the
advantage of making unnecessary the introduction of the 
$T$-matrix approximation to renormalize the order parameter equation, and redefine
the interaction amplitude in terms of the two-body binding energy~\cite{randeria,loktev}.
In the case of 3D, 
$V_{nn} (\mathbf{k}, \mathbf{k'}) \sim 1$ for small $\mathbf{k}$ and 
$\mathbf{k'})$, while 
$V_{nn} (\mathbf{k}, \mathbf{k'}) \sim 1/{k k'}$ for large $\mathbf{k}$ and 
$\mathbf{k'}$, when a generic real potential is used~\cite{nsr}.
Notice that, our three dimensional interaction $\Gamma_n(\mathbf{k}$ has the same behaviour
at both low and high momenta with the one used by Nozieres \textit{et al.}~\cite{nsr} 
$\Gamma(\mathbf{k})=1/\sqrt{1+(k/k_{0})^2}$. Either choice (ours or NSR's)
produces qualitatively similar results as it will be later discussed.

Furthermore, the order parameter equations need to be solved 
self-consistently with the number equation 
\begin{eqnarray}
\mathcal{N}=\sum_{n,\mathbf{k},\sigma}
\left[
\frac{1}{2}-\frac{\xi_n(\mathbf{k})}{2E_n(\mathbf{k})}\tanh\frac{E_n(\mathbf{k})}{2k_bT} 
\right],
\label{numbereqn}
\end{eqnarray}
which is obtained from $\mathcal{N} = -\partial \Omega_0/\partial \mu$, 
where $\Omega_0$ is the saddle point thermodynamic potential.
The inclusion of fluctuations are very important for the number equation 
near the critical temperature of the system, and in this case
the saddle point thermodynamic potential needs to be corrected to 
$\Omega = \Omega_0 + \Omega_G$, where $\Omega_G$ should be calculated at
least at the Gaussian level. In this manuscript, however, we limit ourselves to fluctuation
effects at low temperatures, as discussed next.

\section{Gaussian Fluctuations}
\label{sec:gaussian-fluctuations}

We now investigate Gaussian fluctuations in the pairing field $\Phi_n(q)$
about the the static saddle point $\Delta_{n,0}$.
The Gaussian (quadratic) effective action can be written as 
\begin{equation}
S_{G} = S_{0}(\Delta_{n,0}) + 
\frac{\beta}{2}\sum_q \Lambda^\dagger (-q) \mathbf{M}(q) \Lambda (q)
\end{equation}
where the fluctuation field is
\begin{equation}
\Lambda^\dagger (-q) = \left( \begin{array}{cccc} \Lambda_1^*(q), 
& \Lambda_1(-q), & \Lambda_2(-q), & \Lambda_2^*(q) \end{array}\right)
\end{equation}
and $\mathbf{M} (q)$ is the fluctuation matrix given by
\begin{equation}
\mathbf{M} (q) =\left( \begin{array}{cccc} 
M_{11}^{1}     &  M_{12}^{1}   & 0              & -g_{12} \\ 
M_{21}^{1}     &  M_{22}^{1}   &  -g_{12} &        0      \\ 
     0         & -g_{21} &   M_{22}^{2}   &    M_{21}^{2} \\ 
 -g_{21} &      0        &   M_{12}^{2}   &    M_{11}^{2}
\end{array}\right)
\end{equation}
where the matrix elements $M_{11}^n(q)=M_{22}^n(-q)$, 
and $M_{12}^n(q)=M_{21}^{n*}(q)$ are given by
\begin{eqnarray*}
M_{11}^n&=&-g_{nn} + \beta^{-1}\sum_p |\Gamma_n(p+q/2)|^2
\vartheta_n(p+q)\vartheta_n(-p) \\
M_{12}^n&=& \beta^{-1}\sum_p \Gamma_n^2(p+q/2)
\varsigma_n(p+q)\varsigma_n(p).
\end{eqnarray*}
The expressions 
$
\vartheta_n(p) = 
\bar{\xi}_n(-p)/F(p)
$
and 
$
\varsigma_n(p) = 
\Delta_n(\mathbf{k})/F(p)
$
are the matrix elements $({\mathbf G_{n,0}})_{11}$
and $({\mathbf G_{n,0}})_{12}$, respectively. 
Here, we use the definitions $\bar{\xi}_n(p)=iw_{\ell}-\xi_n(\mathbf{k})$,
and $F(p) = |\bar{\xi}_n(p)|^2+|\Delta_n(\mathbf{k})|^2$.
Notice that while $M_{12}^n(q)$ and $M_{21}^n(q)$ are even under the transformations
$\mathbf{q}\rightarrow -\mathbf{q}$ and $iv_{\ell}\rightarrow -iv_{\ell}$; 
$M_{11}^n(q)$ and $M_{11}^n(q)$ are even 
only under the transformation $\mathbf{q}\rightarrow -\mathbf{q}$, having
no defined parity in $w_{\ell}$. In addition, notice that 
when $g_{12} = g_{21} = 0$,
and the fluctuation matrix becomes block diagonal indicating that the two bands
are uncoupled. This corresponds to the case where the Josephson coupling 
$V_{12} = V_{21} = 0$, since $g_{12} = -V_{12}/\det{\mathbf V}$ as indicated
in Eq.~(\ref{inverseV}).

Performing Matsubara summations over $w_{\ell}$ in the expressions  
for $M_{11}^n$ and $M_{12}^n$ leads to
\begin{eqnarray}
\label{eqn:theta-def}
M_{11}^n(q) &=& -g_{nn}+\Theta_{n,11}^{\rm{qp-qh}} + \Theta_{n,11}^{\rm{qp-qp}}, \\
M_{12}^n(q) &=& \Theta_{n,12}^{\rm{qp-qh}} + \Theta_{n,12}^{\rm{qp-qp}}, 
\end{eqnarray}
where the explicit form of the $\Theta_{n,ij}$ functions is given in Appendix A, 
for the peruse of the reader. We choose to separate the contributions of the
matrix elements $M_{ij}$ in terms of quasiparticle-quasiparticle ($\rm{qp-qp}$) 
and quasiparticle-quasihole ($\rm{qp-qh}$) processes in order to isolate 
the channels that contribute to Landau damping of the collective modes to be
discussed in the next section.

\section{Collective Modes at $T=0$}
\label{sec:collective-modes}

The collective modes are determined by the poles of the propagator 
matrix ${\bf M}^{-1} (q)$ for the pair fluctuation fields 
$\Lambda (q)$, which describe
the Gaussian deviations about the saddle point order parameter. 
The poles of ${\bf M}^{-1} (q)$ are determined
by the condition $\det \mathbf{M}=0$, and lead to a dispersion for the
collective modes $w=w(\mathbf{q})$, when the usual analytic continuation
$iv_{\ell}\rightarrow w+i0^+$ is performed.

We will focus here only at the zero temperature limit, but we will analyse
phase and amplitude modes. At $T=0$, only the $\rm{qp-qp}$ terms contribute, as
the $\rm{qp-qh}$ terms vanish identically (see Appendix A).
In this limit, we separate the diagonal matrix 
elements of ${\mathbf M} (q)$ into even and odd contributions 
with respect to $w$
\begin{eqnarray}
M_{11}^{n,E}(q)&=&-g_{nn}+\sum_{\mathbf{k}}
\frac{\Gamma '^2[\xi\xi '+EE'][E+E']}   {2EE'[ w^2-(E+E')^2]} ,\\
M_{11}^{n,O}(q)&=&-\sum_{\mathbf{k}}
\frac{\Gamma '^2[\xi\xi '+EE'] w}  {2EE'[ w^2-(E+E')^2 ]}.
\end{eqnarray}
The off-diagonal term is even in $w$, and it reduces to
\begin{eqnarray}
M_{12}^n(q)=-\sum_{\mathbf{k}}\frac{\Gamma '^2 \Delta\Delta '[E+E']}  {2EE'[w^2-(E+E')^2]}.
\end{eqnarray}
We used in the previous expressions the following simplified notation,  
the kinetic energies $\xi_n(\mathbf{k}) \to \xi$,
$\xi_n(\mathbf{k}+\mathbf{q}) \to \xi'$;  
the quasiparticle energies $E_n(\mathbf{k}) \to E$,
$E_n(\mathbf{k}+\mathbf{q}) \to E'$; 
the order parameters $\Delta_n(\mathbf{k}) \to \Delta$,
$\Delta_n(\mathbf{k}+\mathbf{q}) \to \Delta'$;
and the symmetry factors 
$\Gamma_n(\mathbf{k}) \to \Gamma$,
$\Gamma_n(\mathbf{k}+\mathbf{q}/2) \to \Gamma'$.

In order to obtain the collective mode spectrum, 
we express $\Lambda_n(q) = \tau_n(q)e^{i\phi_n(q)} = 
( \lambda_n(q) +i \theta_n(q) )/\sqrt{2}$
where $\tau_n(q),\phi_n(q),\lambda_n(q)$ and $\theta_n(q)$ are all real.
Notice that the new fields $\lambda_n (q) = \tau_n (q) \cos \phi(q)$, 
and $\theta_n (q) = \tau_n (q) \sin \phi (q)$ can be regarded essentially
as the amplitude and phase fields respectively, when $\phi(q)$ is small.
This change of basis can be described by the following unitary
transformation 
\begin{eqnarray}
\Lambda (q) =\frac{1}{\sqrt{2}}  \left( \begin{array}{cccc} 
1 & i & 0 & 0 \\
1 &-i & 0 & 0 \\
0 & 0 &-i & 1 \\
0 & 0 & i & 1 
\end{array}\right)   
\left( \begin{array}{c} \lambda_1(q) \\ \theta_1(q) \\ 
\theta_2(q) \\ \lambda_2(q)  \end{array}\right).
\end{eqnarray}
Since we are considering the case where the saddle point order parameters 
$\Delta_1(\mathbf{k})$ and $\Delta_2(\mathbf{k})$ are in phase 
the fluctuation matrix in the rotated basis then reads
\begin{eqnarray}
\label{eqn:fluctuation-matrix}
\mathbf{\widetilde{M}}=
\left( \begin{array}{cccc}
      x_1            & iz_1           &           0             &     -g_{12}      \\
-iz_1         &        y_1            &      -g_{12}      &            0           \\
        0              &      -g_{21}      &         y_2           &    -iz_2      \\
    -g_{21}      &           0             &       iz_2     &	      x_2 
\end{array}\right),
\end{eqnarray}
where $x_n=M_{11}^{n,E}+M_{12}^n$, $y_n=M_{11}^{n,E}-M_{12}^n$, and $z_n=M_{11}^{1,O}$ 
with the $q$ dependence being implicit.
The rotated matrix $\mathbf{\widetilde M}$ is as complex as 
the un-rotated matrix $\mathbf{M}$, and the real advantage
of working in the rotated basis involving fields $\lambda_n$ and
$\theta_n$ is that the interpretation in terms of amplitude and 
phase fields is straight-forward.
For instance, by inspection, notice that the {\it phase}
fluctuation fields $\theta_1$ and $\theta_2$
corresponding to different condensates that
are coupled by interband elements $g_{12}= g_{21}$.
The same applies to {\it amplitude} fluctuation fields 
$\lambda_1$ and $\lambda_2$. When the interaction $V_{12} = 0$, the matrix
element $g_{12}$ vanishes, the matrix $\mathbf {\widetilde M}$ becomes 
block diagonal and the two bands are decoupled.
In this case the one-band results previously obtained are 
recovered for each independent band~\cite{jan}.

Next, we focus on phase-phase and amplitude-amplitude collective modes.
The easiest way to get the amplitude-amplitude collective modes is to integrate out the 
phase fields to obtain an amplitude-only effective action
\begin{equation}
S_{\lambda_1 \lambda_2} = {\beta \over 2} \sum_{q} \lambda^\dagger (q) 
{\mathbf {M_a}}(q) \lambda (q),
\end{equation}
where $\lambda^\dagger (q) = (\lambda_1 (q), \lambda_2 (q))$. 
The amplitude-amplitude fluctuation matrix has the form
\begin{eqnarray}
\mathbf{M_a} = \left( \begin{array}{cc} x_1+y_2z_1^2/W_a  & -g_{12}(1-z_1z_2/W_a) \\ 
-g_{12}(1-z_1z_2/W_a)  & x_2+y_1z_2^2/W_a  \end{array}\right),
\end{eqnarray}
where $W_a=g_{12}^2-y_1y_2$.
The dispersion relation for the amplitude-amplitude collective modes is obtained
from the condition $\det \mathbf{M_a} = 0$. In this paper, however, we are mostly
interested in the phase-phase collective modes.
Thus, upon integration of the amplitude fields we obtain
a phase-only effective action
\begin{equation}
S_{\theta_1 \theta_2} = {\beta \over 2} \sum_{q} \theta^\dagger (q) 
{\mathbf {M_p}}(q) \theta (q),
\end{equation}
where $\theta^\dagger (q) = (\theta_1 (q), \theta_2 (q))$. 
The phase-phase fluctuation matrix has the form
\begin{eqnarray}
\label{eqn:mp}
\mathbf{M_p}=
\left( \begin{array}{cc} y_1+x_2z_1^2/W_p  & -g_{12}(1-z_1z_2/W_p) 
\\ -g_{12}(1-z_1z_2/W_p)  & y_2+x_1z_2^2/W_p  \end{array}\right)
\end{eqnarray}
with $W_p=g_{12}^2-x_1x_2$.
Again, the dispersion relation for the phase-phase collective modes is obtained
from the condition $\det \mathbf{M_p} = 0$ corresponding to poles of the
phase-phase correlation matrix $\mathbf{M_p}^{-1}$. 
In the next section, we discuss both analytical results 
for the phase-phase modes in BCS and BEC limits, as well as
numerical results and the crossover regime. 

We note in passing that at finite temperatures, for continuum $s$-wave systems,
the $\rm{qp-qp}$ terms are well behaved in the long-wavelength and low 
frequency limit, while the $\rm{qp-qh}$ terms do not in general allow 
for a simple expansion in the same limit due to the presence of Landau damping
(see Appendix A). However, at zero temperature, the $\rm{qp-qh}$ terms vanish, 
and a well defined expansion is possible at low frequencies
provided that the collective modes can not decay into the two-quasiparticle continuum.
Thus, the collective mode dispersion $w ({\mathbf q}) $ must satisfy the following
condition
\begin{equation}
w (\mathbf {q}) \ll \min \{E_1(\mathbf{k})+ E_1(\mathbf{k+q}), 
E_2(\mathbf{k})+E_2(\mathbf{k+q})\}.
\end{equation}
To obtain the long wavelength dispersions
for the collective modes at $T = 0$, we expand the matrix elements of 
$\mathbf{\widetilde{M}}$ 
(Eq.~(\ref{eqn:fluctuation-matrix})) in the amplitude-phase representation 
to second order in $\vert {\mathbf q} \vert$ and fourth order in $w$ to get
\begin{eqnarray}
\label{eqn:matrix-elements-second}
x_n&=&A_n + C_n|\mathbf{q}|^2 -D_nw^2 + E_nw^2|\mathbf{q}|^2 + F_nw^4, \\
y_n&=&P_n+Q_n|\mathbf{q}|^2-R_nw^2 + S_nw^2|\mathbf{q}|^2 + T_nw^4, \\
z_n&=&B_nw + H_nw^3,
\end{eqnarray}
where expansion coefficients are given in Appendix B.
As it will become clear in section~\ref{sec:analytical-results}, 
the $|\mathbf{q}|^{4}$ order terms in the expansion are not necessary to calculate
the collective mode frequencies $w(\mathbf{q})$ accurately to order $|\mathbf{q}|^2$.

The expressions of the coefficients found in Appendix B are valid for values of
the interaction range parameter $k_{n,0}$ that satisfy the 
diluteness condition ($k_{n,0}/k_F \gg 1$). 
However, in order to make analytical progress in the calculation of collective
modes to be discussed next, we take the limit 
$k_{n,0}\to\infty$, since all the momentum integrals are convergent.
This is in contrast to the situation encountered with the order parameter
equation, where we used $k_{n,0} \sim 10^4k_F$ in order to ensure convergence
of our numerical calculations for $|\Delta_n|$ and $\mu$ as will be seen in 
the next section.

\section{Analytical and Numerical Results}
\label{sec:analytical-results}

In this section, we will focus only on the long-wavelength (small 
$\vert \mathbf q \vert$) limit phase-phase modes
determined by the condition $\det \mathbf{M_p} = 0$. 
We begin our discussion with the trivial case when $g_{12}=0$ (where $V_{12} = 0$
but $V_{11}$ and $V_{22}$ can have any negative value) 
which corresponds to two uncoupled bands. 
In this case, the phase-phase fluctuation matrix $\mathbf{M_p} = 0$ 
becomes block diagonal, and we find two Goldstone modes satisfying the relation
\begin{equation}
w_n^2 = c_n^2 \vert \mathbf q \vert^2, 
\end{equation}
where the square of the speed of sound $c_n$ in $n^{th}$ band is given by
\begin{equation}
c_n^2 = \frac{ A_nQ_n}{ B_n^2 + A_nR_n}.
\end{equation}
In the limit $\mathbf{q} \to 0$, the corresponding eigenvectors are given by
\begin{equation}
\theta^\dagger (\mathbf{q}=0, w=0) = (\theta_1, \theta_2) \propto (1, 0)
\end{equation}
for the first band, and
\begin{equation}
\theta^\dagger (\mathbf{q}=0, w=0) = (\theta_1, \theta_2) \propto (0, 1)
\end{equation}
for the second band, respectively. These results are identical to the 
known results in the one-band case~\cite{jan}.

However, in the non-trivial case when $g_{12} \ne 0$, which corresponds to a
finite Jopephson coupling $V_{12}$ between bands, we find two modes. 
In order to determine the collective mode spectra $w ({\mathbf q})$
accurately to order $|\mathbf {q}|^2$ it is sufficient to 
rewrite the determinant condition in the form
\begin{equation}
w^4(\alpha_1+\alpha_2|\mathbf{q}|^2) 
+ w^2(\alpha_3+\alpha_4|\mathbf{q}|^2) 
+ \alpha_5|\mathbf{q}|^2=0,
\end{equation}
where $\alpha_n$ are non-trivial and extremely complicated 
functions of the expansion coefficients $A_n, B_n, C_n, D_n, ...$ 
given in Appendix B. The exact dependence can be obtained 
via a symbolic manipulation program, but we will
not quote these general results here or in appendices, as they are not
particularly illuminating. Instead, we will present simple limits,
where their behavior can be easily understood.

In this non-trivial case of $g_{12} \ne 0$, we find two collective modes.
The first mode is the Goldstone mode satisfying the relation
\begin{equation}
w^2 = c^2 \vert \mathbf q \vert^2, 
\end{equation}
where the square of the speed of sound $c$ is given by
\begin{equation}
c^2=-\alpha_5/\alpha_3 > 0 .
\end{equation}
In the limit $\mathbf{q} \to 0$, the eigenvector of the Goldstone mode 
is given by 
\begin{equation}
\label{eqn:eigenvectorg}
\theta^\dagger (\mathbf{q}=0, w=0) = (\theta_1, \theta_2)
\propto (|\Delta_1|,|\Delta_2|)
\end{equation}
which is valid for all values of the $V_{11}$ and $V_{22}$ couplings. 
Notice that this mode is associated 
with the \textit{in-phase} fluctuations of the phases of the
order parameters around their saddle point values. 

In the particular case, where $g_{12}/\min\{N_1,N_2\}$
is the {\it smallest expansion 
parameter} (small $g_{12}$ limit)
we can perform a Taylor expansion of the $\alpha_{n}$ 
coefficients around $g_{12}=0$, 
and obtain
\begin{equation}
\label{smallglimitc}
c^2= t_1t_2(P_2Q_1 + P_1Q_2) / (P_1t_1 + P_2t_2),
\end{equation}
as the square of speed $c$ of the Goldstone mode.
Here, we introduced a coefficient
\begin{equation}
t_n=(A_n-P_n)/[B_n^2+R_n(A_n-P_n)] > 0,
\end{equation}
which is positive definite since $A_n > P_n$ and $R_n > 0$.
The precise meaning of the {\it smallest expansion parameter}
$g_{12}/\min\{N_1,N_2\}$ 
will be clear in sub-sections A and B, where we discuss 
analytically the weak and strong coupling limits.
The eigenvector in the small $g_{12}$ limit 
is exactly the same as in the case of general $g_{12}$,
since $\theta^\dagger (\mathbf{q}=0, w=0) = (\theta_1, \theta_2)$ 
is independent of the value of $g_{12}$. (See Eq.~(\ref{eqn:eigenvectorg}))

The second mode is a finite frequency mode satisfying the relation
\begin{equation}
w^2 = w_0^2 + v^2 \vert \mathbf q \vert^2, 
\end{equation}
where the square of a finite frequency $w_0$ of the mode is
\begin{equation}
w_0^2=-\alpha_3/\alpha_1 > 0
\end{equation}
and the square of the speed $v$ of the mode is
\begin{equation}
v^2=\alpha_3\alpha_2/\alpha_1^2-\alpha_4/\alpha_1+\alpha_5/\alpha_3.
\end{equation}
The eigenvector for this mode has a 
complicated expression
\begin{equation}
\theta^\dagger (\mathbf{q}, w) = (\theta_1, \theta_2) 
\propto (g_{12}(W_p-z_1z_2), y_1W_p + x_2z_1^2),
\end{equation}
for a general value of $g_{12}$.

In the small $g_{12}$ limit, the coefficients simplify to 
\begin{eqnarray}
\label{smallglimitw0}
w_0^2 & = & P_1t_1 + P_2t_2 > 0, \\
\label{smallglimitv}
v^2   & = & Q_1t_1 + Q_2t_2 - c^2 > 0.
\end{eqnarray}
In the limit of $w=w_0$, $\mathbf{q} \to 0$, and small $g_{12}$, 
the eigenvector expression simplifies to
\begin{equation}
\theta^\dagger (\mathbf{q}, w) = (\theta_1, \theta_2) 
\propto (|\Delta_2|t_1, -|\Delta_1|t_2),
\end{equation}
and it becomes transparent that this mode is associated with \textit{out-of-phase} 
fluctuations of the phases of the order parameters 
around their saddle point values, since $t_n$ are positive definite.

Before discussing the collective modes in the analytically tractable
weak and strong coupling limits, notice that a finite $\mathbf {q}$ 
Goldstone mode is only possible when $c^2 > 0$, and that a finite 
$\mathbf {q}$ finite frequency mode is only possible
when $w_0^2 > 0$. 
If these conditions are violated the modes are non-existent. 
Furthermore, caution should be exercised by recalling that 
the small $w$ approximation
used in the expansion of general fluctuation matrix 
($\mathbf{\widetilde{M}}$) elements breaks down when the 
frequency $w$ of any of the
collective modes moves up into the continuum of two-quasiparticle states.
Therefore, our results are strictly valid only for 
\begin{equation}
w \ll \min \{2E_1(\mathbf{k}),2E_2(\mathbf{k})\},
\end{equation}
which corresponds to 
\begin{equation}
w \ll  \min\{2|\Delta_1|,2|\Delta_2|\}
\end{equation}
in the weak coupling (BCS) limit, and to 
\begin{equation}
w \ll \min \{ 2\sqrt{|\mu|^2+|\Delta_1|^2}, 2\sqrt{(|\mu|+E_0)^2+|\Delta_2|^2} \}
\end{equation}
in the strong coupling (BEC) limit.
With these conditions in mind, we discuss next the weak and strong coupling 
limits.

\subsection{Weak Coupling Limit} 

The $s$-wave weak coupling limit is characterized by
the criteria $\mu > 0$, $\mu > E_0$, $\mu\approx\varepsilon_{F} \gg \vert \Delta_1 \vert$, 
and $\mu-E_0\gg \vert \Delta_2 \vert$. 
Analytic calculations are particularly simple in this case
since all integrals for the coefficients needed to calculate the collective mode
dispersions are peaked near the Fermi surface (see Appendix B).
In addition, we make use of the nearly perfect particle-hole symmetry, which forces 
integrals to vanish, when their integrands are odd under the transformation 
$\xi\rightarrow-\xi$.  For instance, the coefficients that couple phase and amplitude
modes within a given band ($B_n$ and $H_n$) vanish. Thus, in this case, there 
is no mixing between phase and amplitude fields within $n^{th}$ band, 
as can be seen by inspection of the fluctuation matrix $\widetilde\mathbf{M}$. 

We would like to focus on the phase-phase collective modes, 
as they correspond to the low energy
part of the collective mode spectrum. Notice that, 
all expansion coefficients appearing in the phase-phase fluctuation matrix
$\mathbf {M_p}$ are analytically tractable.  
The expansion of the matrix elements to order $|{\mathbf q}|^4$ and $w^4$ is
performed under the condition $(w,|\mathbf{q}|^2/2m_n)\ll \min \{2|\Delta_1|,2|\Delta_2|\}$.
To evaluate $x_n$ for each band $n$, we need 
$A_1=g_{12}|\Delta_2|/|\Delta_1|+N_1$, 
$A_2=g_{21}|\Delta_1|/|\Delta_2|+N_2$,
which are the coefficients of the $({\mathbf q} = 0, w = 0)$ term;
$C_n=c_{n,w}^2N_n/12|\Delta_n|^2$, which are the coefficients of $|\mathbf{q}|^2$;
$D_n=N_n/12|\Delta_n|^2$, 
which are the coefficients of $w^2$;
and $F_n=-N_n/120|\Delta_n|^4$,
which are the coefficients of $w^4$.
To evaluate $y_n$, we need
$P_1=g_{12}|\Delta_2|/|\Delta_1|$ and
$P_2=g_{21}|\Delta_1|/|\Delta_2|$,
which are the coefficients of the  $({\mathbf q} = 0, w = 0)$ term;
$Q_n=c_{n,w}^2N_n/4|\Delta_n|^2$, which are the coefficients of $|\mathbf{q}|^2$;
$R_n=N_n/4|\Delta_n|^2$,
which are the coefficients of $w^2$;
and $T_n=-N_n/24|\Delta_n|^4$
which are the coefficients of $w^4$.
Here, 
\begin{equation}
c_{n,w}=v_{n,F}/\sqrt d_n
\end{equation}
is the velocity of the sound mode in the one-band case~\cite{carlos},
$d_n$ is the dimension, $v_{n,F}$ is the Fermi velocity, and $N_n$ is the density of states 
at the Fermi energy per spin ($N_n = m_n L^2/2\pi $ in 2D and $N_n = m_nL^3k_{n,F}/2\pi^2$ in 3D)
in $n^{th}$ band. 
While we use $N_n$ as the density of states per spin at the Fermi energy,
in Ref.~\cite{leggett} the density of states used includes both spins.
The off-diagonal elements are 
$(\mathbf{M_p})_{12}= (\mathbf{M_p})_{21} = -g_{12}$, 
since $B_1 = B_2 = 0$ and $H_1 = H_2 = 0$, as discussed above. 
Notice that the expressions above are valid for both 2D and 3D bands.

In order to bring our results in contact with Leggett's~\cite{leggett}, and 
Sharapov {\it et. al.}~\cite{sharapov}, we make use of the order parameter saddle
point Eq.~(\ref{orderparameter}) at $T = 0$
\begin{eqnarray}
|\Delta_1|(1+V_{11}F_1)&=&-|\Delta_2|V_{21}F_2, \\
|\Delta_2|(1+V_{22}F_2)&=&-|\Delta_1|V_{12}F_1, 
\end{eqnarray}
and consider the small $g_{12}$ limit where
\begin{equation}
g_{12}/\min\{N_1,N_2\} \ll \min\{|\Delta_1|,|\Delta_2|\}/\max\{|\Delta_1|,|\Delta_2|\}.
\end{equation}
A simple evaluation of $\det \mathbf{M_p} = 0$ leads to a Goldstone mode $w^2=c^2|\mathbf{q}|^2$, 
characterized by the speed of sound
\begin{equation}
\label{weakcouplingc}
c^2=\frac{N_1c_{1,w}^2+N_2c_{2,w}^2}{N_1+N_2}, 
\end{equation}
and a finite frequency (Leggett) mode $w^2=w_0^2+v^2|\mathbf{q}|^2$,
characterized by 
\begin{eqnarray}
\label{eqn:weakcouplingw0}
w_0^2 & = & \frac{N_1+N_2}{2N_1N_2} 
\frac{|8V_{12}||\Delta_1||\Delta_2|}{V_{11}V_{22}-V_{12}^2}, \\
v^2 & = & \frac{N_1c_{2,w}^2 + N_2c_{1,w}^2}{N_1+N_2},
\end{eqnarray}
where $w_0$ is a finite frequency, and $v$ is the speed of propagation 
of the mode. Here we reintroduced all the coupling constants of the original Hamiltonian
in Eq.~(\ref{hamiltonian}). These results are valid only in the weak-coupling limit,
with all $V_{nm} < 0$, and ${\mathbf \det \mathbf{V}} > 0$. 
Notice that if $V_{12} = 0$ the Leggett
mode does not exist as the two bands are uncoupled. Furthermore, the trivial limit of one-band (say only band 1 exists) is directly recovered 
by taking $|\Delta_2| = 0$, $N_2 = 0$ and $c_{2,w}=0$ 
which leads to $c^2=c_{1,w}^2$, $w_0^2=0$, and $v^2=0$.

It is also very illustrative to analyse the eigenvectors associated with 
the two-solutions. For Goldstone's mode, in the limit of ${\mathbf q} \to 0$;
for any $g_{12}$, it is easy to see that 
\begin{equation}
\theta^\dagger ({\mathbf q = 0}, w = 0) = (\theta_1, \theta_2) 
\propto (|\Delta_1|, |\Delta_2|),
\end{equation}
corresponding to an \textit{in-phase} mode. 
In the degenerate case, where $|\Delta_1| = |\Delta_2|$,
and $N_1 = N_2$, the eigenvector 
$\theta^\dagger ({\mathbf q = 0}, w = 0) \propto (1, 1)$ and the phase fields
are perfectly in phase.
The eigenvector for the finite frequency mode is
$\theta^\dagger (\mathbf {q}, w) = (\theta_1, \theta_2) 
\propto (g_{12}, y_1)$.
In the particular limit where $g_{12}$ is small (Leggett mode), 
this leads to an eigenvector
\begin{equation}
\theta^\dagger ({\mathbf q = 0}, w = w_0) = (\theta_1, \theta_2) 
\propto (N_2 |\Delta_1|, - N_1 |\Delta_2|),
\end{equation}
which corresponds to an \textit{out-of-phase} mode. In the degenerate case, 
this simplifies further
to $\theta^\dagger ({\mathbf q = 0}, w = 0) = (1,-1)$ becoming a perfectly 
out-of-phase mode.

Now, we would like to turn our attention to the analysis of 
the phase-phase modes in the strong coupling limit, which 
is also analytically tractable.

\subsection{Strong Coupling Limit}

The $s$-wave strong coupling limit is characterized by
the criteria $\mu < 0$, and $\vert \mu \vert \gg \vert \Delta_1 \vert$ and 
$\vert \mu \vert  + E_0 \gg \vert \Delta_2 \vert$.
The situation encountered here is very different from the weak coupling 
limit, because one can no longer invoke particle-hole symmetry to simplify the
calculation of many of the coefficients appearing in the fluctuation matrix 
$\widetilde \mathbf M$ (see Appendix B).
In particular, the coefficients
$B_n \ne 0$ indicate that the amplitude and phase fields within 
an individual band $n$ are mixed.

We will concentrate here only on phase-phase modes characterized by the 
fluctuation matrix $\mathbf {M_p}$.
The expansion of the matrix elements to order $|{\mathbf q}|^2$ and $w^4$ is
performed under the condition $(w,|\mathbf{q}|^2/2m_n) \ll 2|\mu|$.
All the coefficients $(x_n, y_n, z_n, W_p)$ appearing in $\mathbf{M_p}$ 
matrix Eq.~(\ref{eqn:mp}) are 
evaluated to order $(\vert\Delta_1\vert / \vert\mu\vert)^2$ and 
$(\vert\Delta_2\vert / (\vert \mu \vert + E_0))^2$ in the strong coupling limit
for both two and three dimensional systems.
To evaluate $x_n$ for each band $n$, we need 
$A_1=g_{12}|\Delta_2|/|\Delta_1|+\kappa_1|\Delta_1|^2/2|\mu|$, 
$A_2=g_{21}|\Delta_1|/|\Delta_2|+\kappa_2|\Delta_2|^2/2(|\mu|+E_0)$,
which are the coefficients of the $({\mathbf q} = 0, w = 0)$ term;
$C_n=\kappa_n/4m_n$, which are the coefficients of $|\mathbf{q}|^2$;
$D_1=\kappa_1/8|\mu|$,  
$D_2=\kappa_2/8(|\mu|+E_0)$, 
which are the coefficients of $w^2$,
and
$F_1=-\kappa_1\gamma_1/512|\mu|^3$,
$F_2=-\kappa_2\gamma_1/512(|\mu|+E_0)^3$,
which are the coefficients of $w^4$.
To evaluate $y_n$, we need
$P_1=g_{12}|\Delta_2|/|\Delta_1|$ and
$P_2=g_{21}|\Delta_1|/|\Delta_2|$,
which are the coefficients of the  $({\mathbf q} = 0, w = 0)$ term;
$Q_n=\kappa_n/4m_n$, which are the coefficients of $|\mathbf{q}|^2$;
$R_1=\kappa_1/8|\mu|$,  
$R_2=\kappa_2/8(|\mu|+E_0)$, 
which are the coefficients of $w^2$,
and
$T_1=-\kappa_1\gamma/512|\mu|^3$,
$T_2=-\kappa_2\gamma/512(|\mu|+E_0)^3$,
which are the coefficients of $w^4$.
To evaluate $z_n$ for each band $n$, we need 
$B_n=\kappa_n$, which are the coefficients of $w$,
and
$H_1=\kappa_1\bar\gamma/96|\mu|^2$,
$H_2=\kappa_2\bar\gamma/96(|\mu|+E_0)^2$
which are the coefficients of $w^3$.
The computation of $W_p=g_{12}^2-x_1x_2$, just relies
on the knowledge of $x_n$ already obtained above.
While the expressions above are valid for any value of $\mu$ and $E_0$
in the 2D bands, they are only rigorously valid for $|\mu| \gg E_0$ in the 3D case
through the use of the dimensionally dependent variables $\kappa_n$, 
and $\gamma$ and $\bar{\gamma}$ which in the 3D case become
\begin{eqnarray}
\kappa_1 &=& \pi N_1 /8\sqrt{\vert \mu \vert \varepsilon_{F}}, \\
\kappa_2 &=& \pi N_2 /8\sqrt{(\vert \mu \vert + E_0)\varepsilon_{F}}, 
\end{eqnarray}
and $\gamma=5$, $\bar\gamma=3$. To write our expressions in compact notation
and to recover the degenerate limit ($E_0 \to 0$), 
we keep $E_0$ in all 3D coefficients. 
All expressions for the 3D case can be converted into
the 2D case through the following procedure. 
In all coefficients $(x_n, y_n, z_n, W_p)$, the dependence in 
$|\mu|,(|\mu|+E_0)$ is transformed into $|\mu|/2, (|\mu|+E_0)/2$, and
the variables $\kappa_n$, and $\gamma$ and $\bar{\gamma}$ need to be redefined as
\begin{eqnarray}
\kappa_1 &=& N_1/8\vert \mu \vert, \\
\kappa_2 &=& N_2/8(\vert \mu \vert + E_0),
\end{eqnarray}
and $\gamma=2$, $\bar\gamma=2$. 

In the limit $\mathbf{q}\rightarrow 0$, the condition $\det \mathbf{M_p} = 0$ leads 
again to two modes. General expressions for the collective modes are 
highly non-trivial and, therefore, we consider analytically only the asymptotic
small $g_{12}$ limit defined as 
\begin{equation}
g_{12}/\min\{N_1,N_2\} \ll \left[|\min\{|\Delta_1|,
|\Delta_2|\}|/(|\mu|+E_0)\right]^2 \ll 1. 
\end{equation}
w
In this case, the Goldstone mode corresponds to $w^2=c^2|\mathbf{q}|^2$, where
\begin{equation}
\label{strongcouplingc}
c^2=\frac{\kappa_1|\mu|c_{1,s}^2+\kappa_2(|\mu|+E_0)c_{2,s}^2}{\kappa_1|\mu|+\kappa_2(|\mu|+E_0)} 
\end{equation}
is the square of the speed of sound. 
The finite frequency mode (which is the extension of Leggett's mode in the 
weak coupling limit) corresponds to $w^2=v^2|\mathbf{q}|^2+w_0^2$, where
\begin{eqnarray}
\label{strongcouplingw0}
w_0^2 & = & \frac{\kappa_1|\mu|+\kappa_2(|\mu|+E_0)}{2\kappa_1|\mu|\kappa_2(|\mu|+E_0)} 
\frac{\vert V_{12}\Delta_1\Delta_2 \vert}{V_{11}V_{22}-V_{12}^2}, \\
v^2 & = & \frac{\kappa_1|\mu|c_{2,s}^2 + \kappa_2(|\mu|+E_0)c_{1,s}^2}{\kappa_1|\mu|+\kappa_2(|\mu|+E_0)},
\end{eqnarray}
are the finite frequency, and speed of propagation of the mode, respectively. 
Here, the quantitites
\begin{eqnarray} 
c_{1,s} &=& |\Delta_1|/\sqrt{8m_1|\mu|}, \\
c_{2,s} &=& |\Delta_2|/\sqrt{8m_2|(\mu|+E_0)}
\end{eqnarray}
are the velocities of the sound mode in the one-band case~\cite{jan}.
These results are valid only in the strong-coupling limit,
with all $V_{nm} < 0$, and ${\mathbf \det \mathbf{V}} > 0$. 
Notice that if $V_{12} = 0$ the finite frequency mode does not
exist as the two bands are uncoupled. Furthermore, the trivial limit of one
band (say only band 1 exists) is directly recovered by taking $\kappa_2 = 0$ 
and $c_{2,s}=0$ which leads to $c^2=c_{1,s}^2$, $w_0^2=0$, and $v^2=0$.

The eigenvectors associated with these solutions are as follows. 
For Goldstone's mode, in the limit of ${\mathbf q} \to 0$ and 
for any value of $g_{12}$,
\begin{equation}
\theta^\dagger ({\mathbf q = 0}, w = 0) = (\theta_1, \theta_2) 
\propto (|\Delta_1|, |\Delta_2|),
\end{equation}
corresponding to an \textit{in-phase} mode. 
For the finite frequency mode (in the small $g_{12}$ limit), the eigenvector becomes
\begin{equation}
\theta^\dagger ({\mathbf q = 0}, w = w_0) 
\propto (\kappa_2(|\mu|+E_0)|\Delta_1|, - \kappa_1|\mu||\Delta_2|),
\end{equation}
which corresponds to an \textit{out-of-phase} mode, as $\kappa_n > 0$.
In the degenerate case ($\kappa_1 = \kappa_2$, $|\Delta_1| = |\Delta_2|$ and
$E_0 \to 0$), this simplifies further
to $\theta^\dagger ({\mathbf q = 0}, w = 0) = (1,-1)$ 
becoming a perfectly out-of-phase mode, similar to the weak coupling case.

Next, we would like to turn our attention 
to the analysis of the phase-phase modes in the
crossover region which is not analytically tractable 
and requires numerical calculations.

\subsection{Numerical Results and Crossover Region}

Thus far, we have focused on the analytically tractable limits
corresponding to weak and strong couplings. In order to gain further
insight into the behavior of the phase-phase collective excitations 
at $T = 0$, we present numerical results of the evolution from weak to strong 
coupling for the Goldstone and finite frequency modes. 
We limit ourselves to numerical calculations of the fully degenerate 
(identical) bands case, from which the known one-band results~\cite{jan} 
for the Goldstone mode can be easily recovered in the BCS, BEC, 
and crossover regimes for an $s$-wave superfluid.  
While the limit of degenerate bands is probably harder to find in nature, 
it provides us with qualitative and quantitative understanding 
of the evolution of the finite frequency collective modes from weak 
to strong coupling. At the same time the degenerate problem 
is easier to solve numerically, and serves as a test model to our analytical 
results. Thus, we postpone a detailed numerical calculation of non-denegerate 
bands case for a future publication, where the more complex numerical problem
will be attacked.

\begin{figure*} [ht]
\centerline{\scalebox{0.55}{\includegraphics{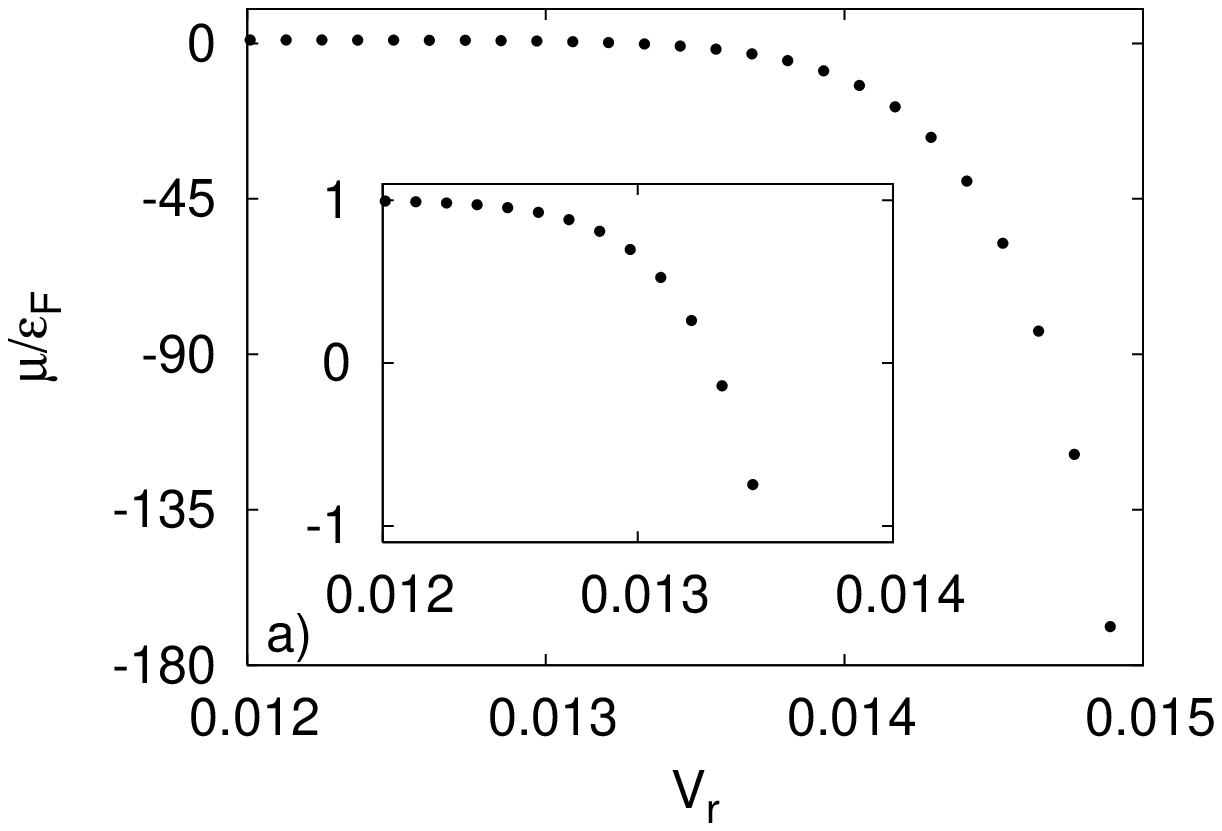}}
\scalebox{0.55}{\includegraphics{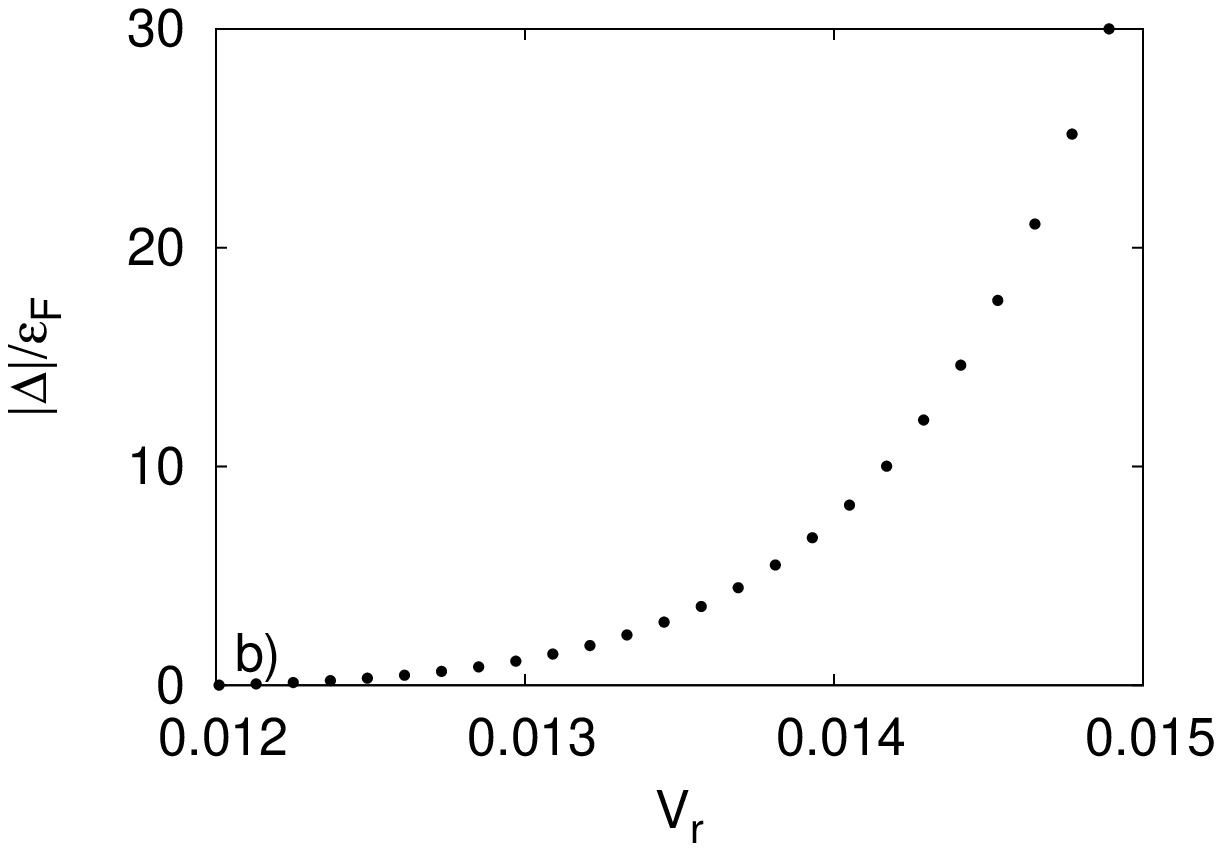}}}
\caption{\label{2D.mu.gap} Plots of a) chemical potential $\mu$ scaled
by $\varepsilon_F$
and b) order parameter $|\Delta|$ scaled by $\varepsilon_F$ 
versus dimensionless coupling $V_r$ (see text for definition) 
for two-dimensional degenerate bands. The region where $\mu$ changes 
sign is shown in the inset. Note that $\mu=0$ when $V_r=0.0133$.}
\end{figure*}
\begin{figure*} [ht]
\centerline{\scalebox{0.55}{\includegraphics{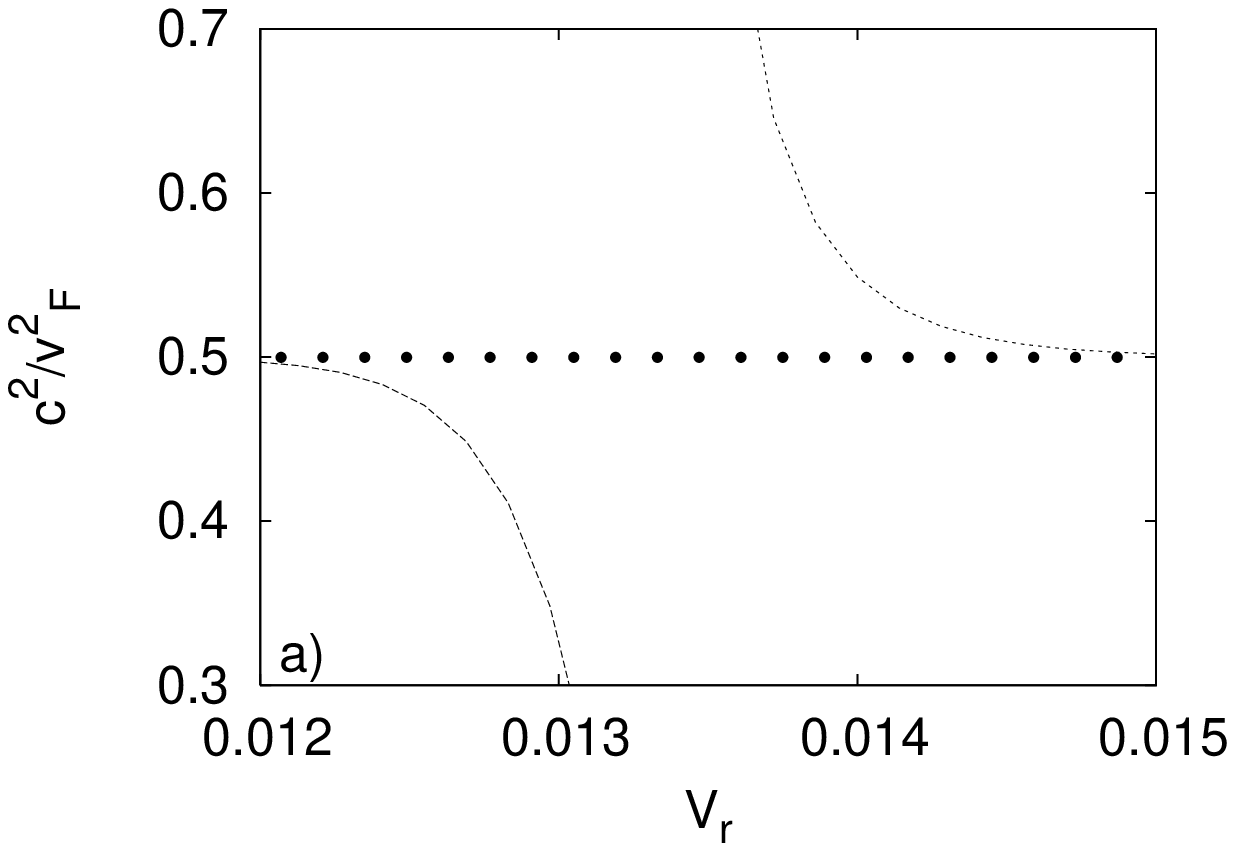}}
\scalebox{0.55}{\includegraphics{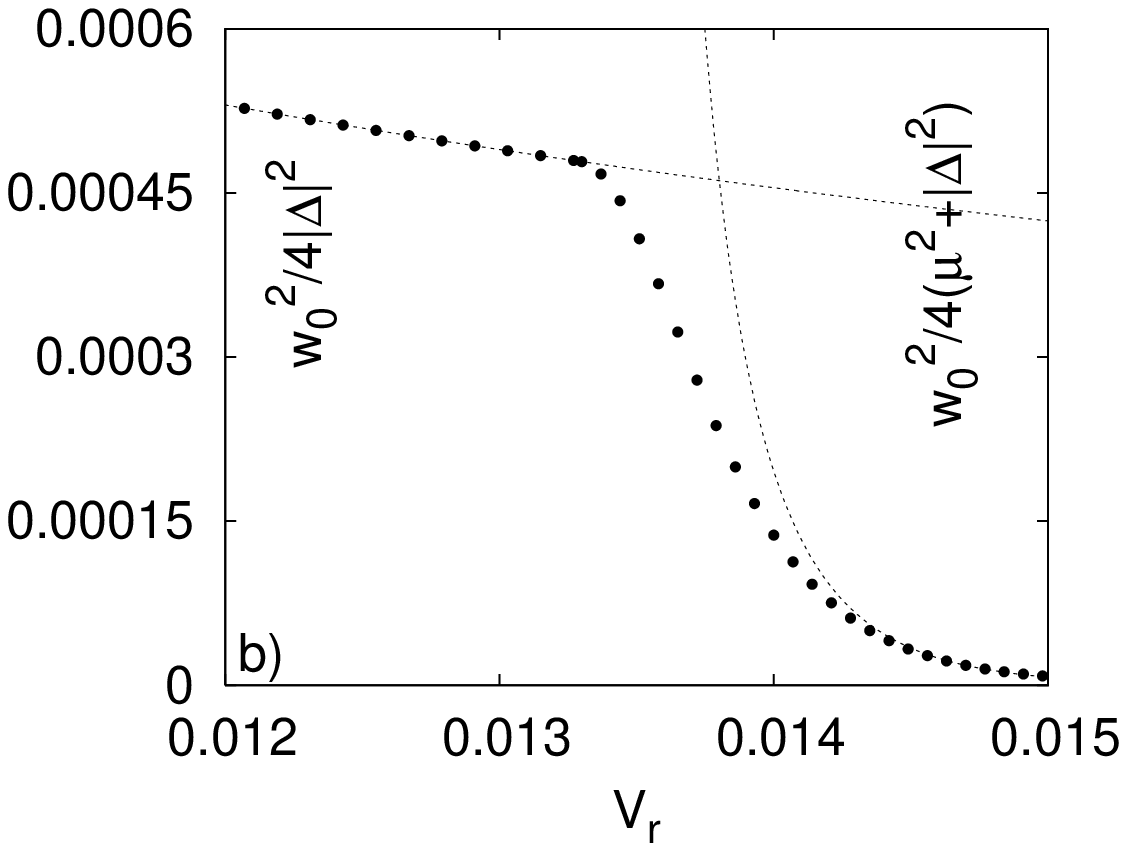}}}
\caption{\label{2D.w0.s} Numerical calculation (solid circles) of
a) square of sound velocity $c^2$ scaled by $v_F^2$,
and of b) the square of finite frequency $w_0^2$ scaled
by the two-quasiparticle threshold $\min \{2 E({\mathbf k}) \}$
versus dimensionless coupling $V_r$
for two-dimensional degenerate bands.
The dotted lines represent analytical results for the
weak and strong coupling limits. 
Notice that, we scaled $w_0^2$ by $4|\Delta|^2$ for $\mu > 0$, and
by $4(|\mu|^2 + |\Delta|^2)$ for $\mu < 0$. }
\end{figure*}

In this particular case (identical bands), 
the band offset is $E_0 \to 0$, the density of states at the Fermi energy
are $N_1 = N_2 = N$; the Fermion masses are $m_1 = m_2 = m$; 
the Fermi velocities are $v_{1,F}= v_{2,F} = v_F$; 
and the intra-band interactions are
$V_{11} = V_{22} = V$; while the order parameter amplitudes are
$|\Delta_1| = |\Delta_2| = |\Delta|$).
Furthermore, the low-momentum and low-energy expansion 
coefficients become identical, i.e.,
$A_1 = A_2 = A$; $B_1 = B_2 = B$; $D_1 = D_2 = D$; 
$H_1 = H_2 = H$; $R_1 = R_2 = R$; and $T_1 = T_2 = T$, 
and we define 
$\tilde{A} = A-g_{12} = \sum_{\mathbf{k}}\Delta(\mathbf{k})^2/2E(\mathbf{k})^3$. 
For any value of $g_{12}$ in the degenerate bands limit, we obtain 
\begin{equation}
\label{degeneratec}
c^2 = v^2 = \frac
{ Q\tilde{A}  }
{ B^2 + \tilde{A}R } > 0, 
\end{equation}
for the squares of the speeds $c$ (Goldstone)
and $v$ (finite frequency). 
In addition, we obtain
\begin{equation}
\label{degeneratew0}
w_0^2 = \frac{\mathcal{P}_2(g_{12})} {\mathcal{P}_2'(g_{12})}
\end{equation}
for the square of the finite frequency which is also valid 
for any value of $g_{12}$. The numerator
$ \mathcal{P}_2(g_{12}) = 2(B^2 + \tilde{A}R) (g_{12}\tilde{A} + 2g_{12}^2) $
is positive definite and the denominator
$ \mathcal{P}_2'(g_{12}) = (B^2 + \tilde{A}R)^2 
+ g_{12}[ R(B^2 + \tilde{A}R) + \tilde{A}(2T\tilde{A}-4BH) + 8\tilde{A}RD + 6B^2D ]
+ g_{12}^2 [4T\tilde{A} + 8(RD-BH) + 5B^2D/\tilde{A}] $
must be also positive definite to guarantee that $w_0^2 > 0$. 
Notice that these functions are not strictly second order 
polynomials in $g_{12}$, since all coefficients 
depend implicitly on $g_{12}$.
In the small $g_{12}$ limit, expressions for $c^2$, $v^2$ and $w_0^2$ 
simplify to Eqs.~(\ref{smallglimitc}), ~(\ref{smallglimitv}) 
and ~(\ref{smallglimitw0}) with the degenerate case coefficients. 
Furthermore, since $\mathcal{P}_2(g_{12}) > 0$ is positive definite, 
Eq. (\ref{degeneratew0}) is only valid strictly for $\mathcal{P}_2'(g_{12}) > 0.$

\begin{figure*} [ht]
\centerline{\scalebox{0.55}{\includegraphics{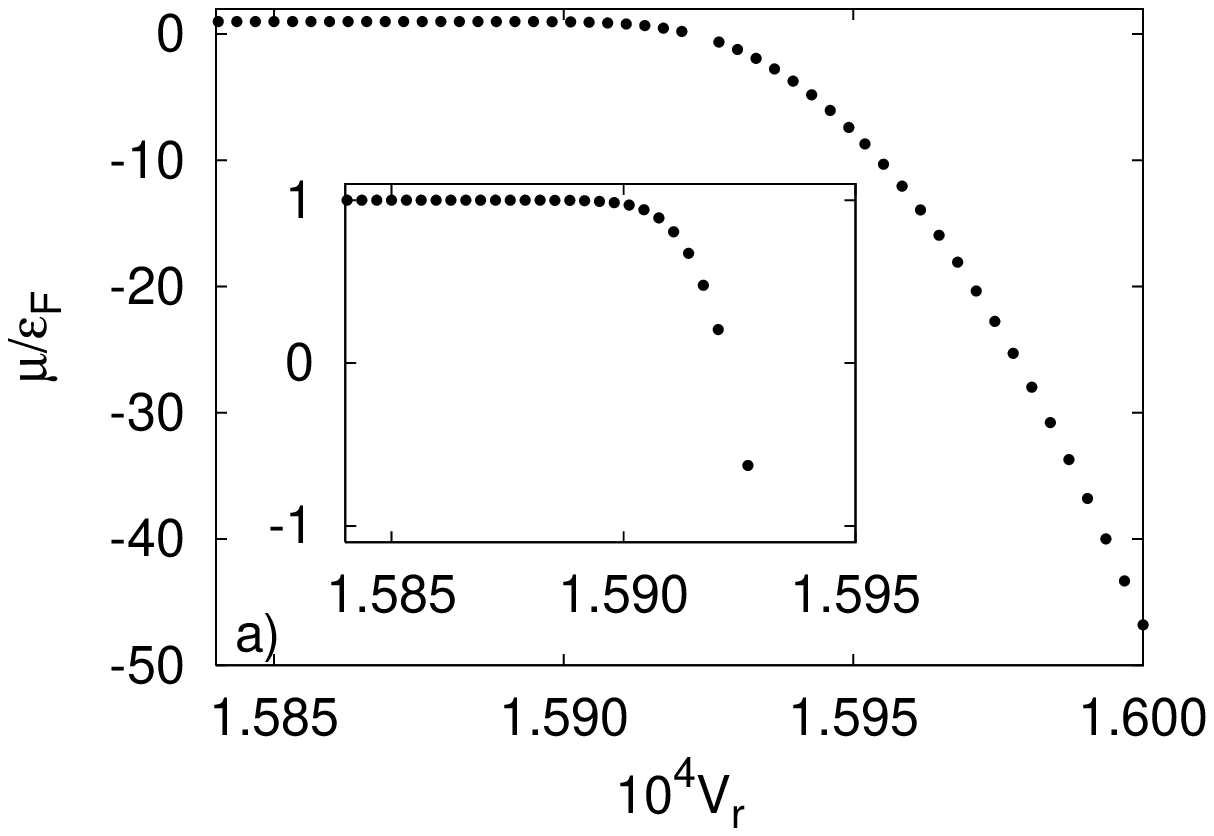}}
\scalebox{0.55}{\includegraphics{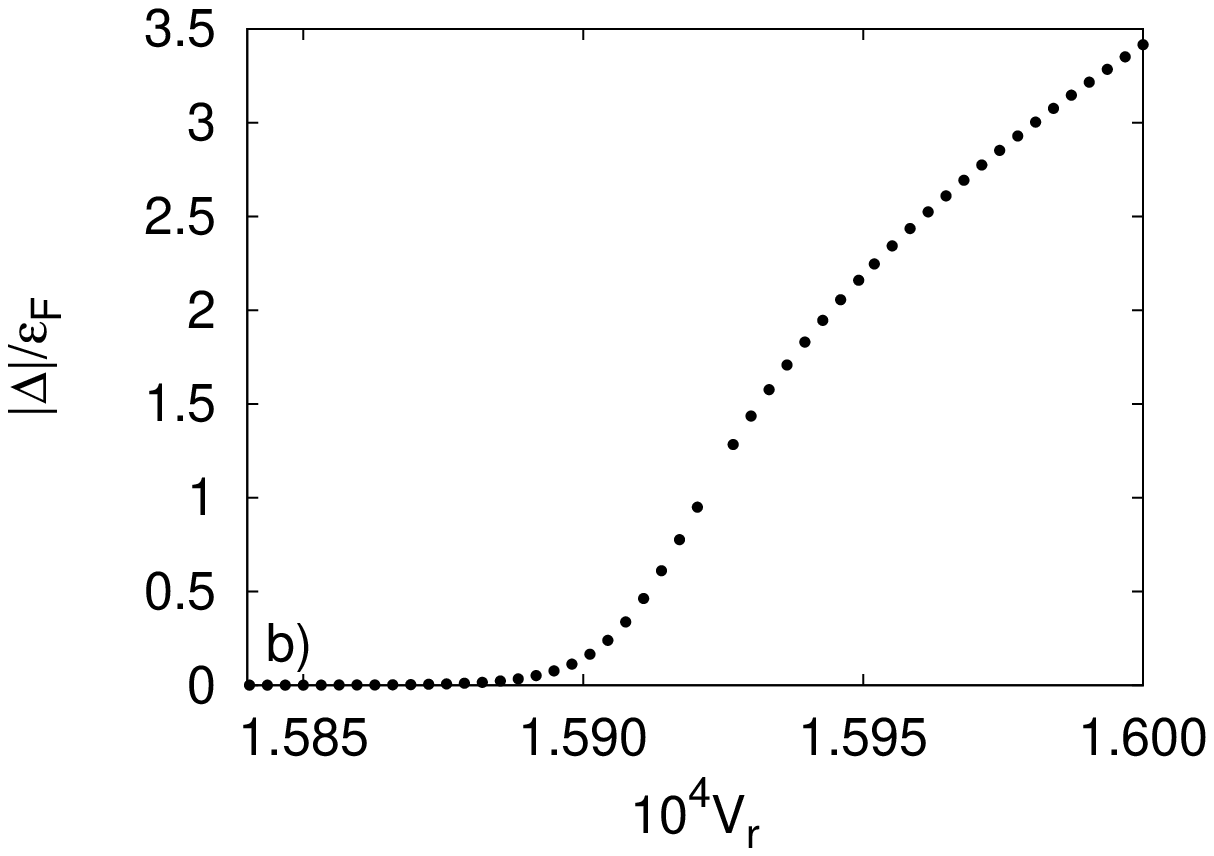}}}
\caption{\label{3D.mu.gap}  Plots of a) chemical potential $\mu$ scaled
by $\varepsilon_F$
and b) order parameter $|\Delta|$ scaled by $\varepsilon_F$ 
versus dimensionless coupling $V_r$ (see text for definition) 
for three-dimensional degenerate bands. The region where $\mu$ changes 
sign is shown in the inset. Note that $\mu=0$ when $V_r=1.592 \times 10^{-4}$.}
\end{figure*}
\begin{figure*} [ht]
\centerline{\scalebox{0.55}{\includegraphics{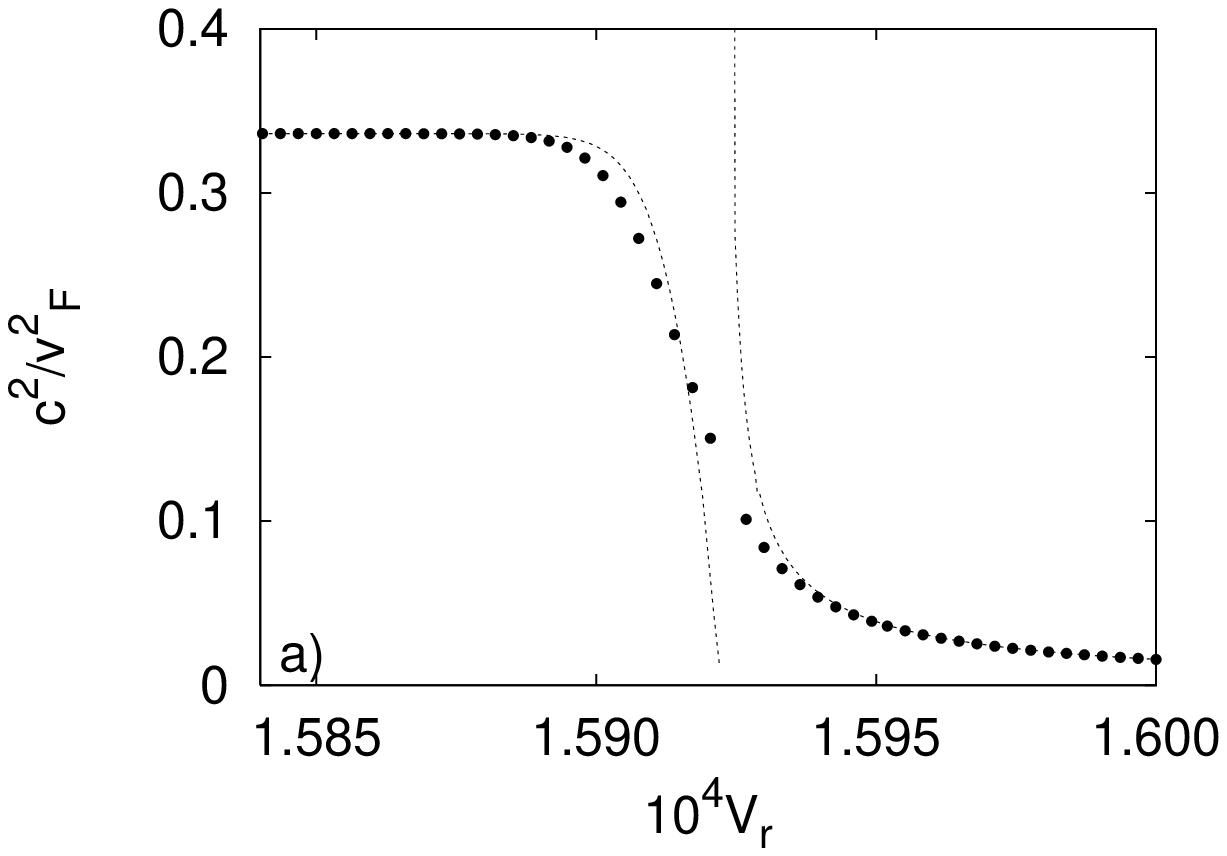}}
\scalebox{0.55}{\includegraphics{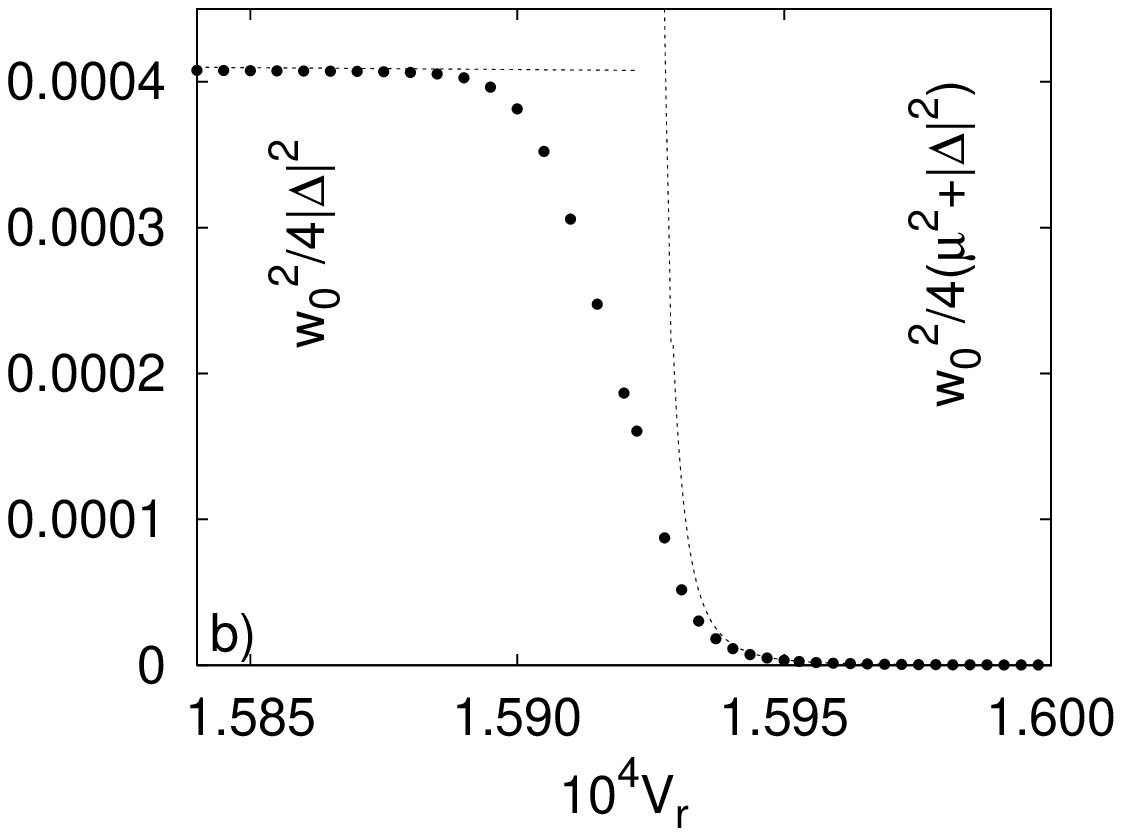}}}
\caption{\label{3D.w0.s} 
Numerical calculation (solid circles) of
a) square of sound velocity $c^2$ scaled by $v_F^2$,
and of b) the square of finite frequency $w_0^2$ scaled
by the two-quasiparticle threshold $\min \{2 E({\mathbf k}) \}$
versus dimensionless coupling $V_r$
for three-dimensional degenerate bands.
The dotted lines represent analytical results for the
weak and strong coupling limits. 
Notice that, we scaled $w_0^2$ by $4|\Delta|^2$ for $\mu > 0$, and
by $4(|\mu|^2 + |\Delta|^2)$ for $\mu < 0$. }
\end{figure*}

In our numerical calculations of $|\Delta_n|$ and $\mu$
(via the saddle point Eqs.~(\ref{orderparameter}) and~(\ref{numbereqn})),
we choose a momentum cut-off of value 
$k_{1,0} = k_{2,0} = k_0 \sim 10^4k_F$ 
to ensure convergenge of all $\mathbf{k}$-space integrations. 
The fermion density can be expressed in dimensionless units
as $n_s = n/n_{max}$, 
where $n_{max} = {k_F}_{max}^2/2\pi$ in 2D, and 
$n_{max} = {k_F}_{max}^3/3\pi^2$
in 3D. The maximal value of the Fermi momentum ${k_F}_{max}$
(that fixes the maximal density $n_{max}$)
is chosen by fixing the ratio ${k_F}_{max} / k_0 = 10^{-4}$, 
which easily satisfies the diluteness 
conditions $(k_0/ {k_F}_{max})^3 \gg 1$ (or $n_{max} R_0^3 \ll 1$) in 3D,
and $(k_0/ {k_F}_{max})^2 \gg 1$ (or $n_{max} R_0^2 \ll 1$) in 2D.
See Sec.~\ref{sec:effective-action-method} to recall how the interactions
depend on $k_0 = 2\pi/R_0$, where $R_0$ plays the role of the interaction
range in real space. For any $k_F/{k_F}_{max} < 1$ all conditions are satisfied,
thus we work at fixed density $n_s = 1/2\pi \approx 0.159$, corresponding to 
$k_F/{k_F}_{max} = 1/\sqrt{2\pi} = 0.398$ in 2D, and 
$n_s = 3/(8\pi) \approx 0.119 $, corresponding
to $k_F/{k_F}_{max}= (3/8\pi)^{1/3} \approx 0.492$ in 3D.

We also confine ourselves to the asymptotic small $g_{12}$ limit, which
means $g_{12}/N \ll 1$ in weak coupling, and $g_{12}/N \ll [|\Delta|/|\mu|]^2$ 
in strong coupling limits. Therefore, in 3D, we choose $V_{12} = 10^{-7} V$ (since 
$V_{11} = V_{22} = V$) which leads to $g_{12}/N \sim 10^{-4}$,
for a 3D Fermion density of $n_s = 0.119$. In the 2D case, we choose
$V_{12} = 10^{-5} V$, which leads to $g_{12}/N \sim 10^{-4}$, for
a 2D Fermion density of $n_s = 0.159$. 
This particular choice satisfies the small $g_{12}$ condition 
for the range of couplings shown in Figs.~2 and ~4. 

We solve the saddle point equations for the order parameter 
Eq.~(\ref{orderparameter}) together with the number equation Eq.~(\ref{numbereqn})
self-consistently for fixed densities.
The order parameter amplitude $|\Delta|$ and chemical potential 
$\mu$ in 2D (3D) are presented in Figs.~2a and~2b
(Figs.~4a and~4b) as a function of the dimensionless 
intra-band interaction parameter $V_r = NV/\pi$. 
Notice that the system crosses over from the BCS ($\mu>0$) to BEC ($\mu<0$) regimes
at $V_r=1.33 \times 10^{-2}$ in 2D,
and at $V_r = 1.59 \times 10^{-4}$ in 3D, where the chemical
potential $\mu$ crosses zero (the bottom of the degenerate bands).

For the 2D (3D) case, we show in Figs.~3a and~3b (Figs.~5a and~5b),
numerical plots of the sound velocity $c$,
normalized by the Fermi velocity $v_{F}$, 
and of the ratio of the finite frequency $w_0$ 
with respect to the minimum quasi-particle excitation energy 
$\min \{2E(\mathbf{k})\}$ as a function of intra-band couplings $V_r$. 
In addition, notice the very good agreement between the numerical results and 
the analytical approximations in their respective (BCS or BEC) limits.

The analytical value for the weak coupling sound velocity follows 
from Eq.~(\ref{weakcouplingc}) for the non-degenerate case 
with $N_1 = N_2$ and $c_{1,w} = c_{2,w}$, 
which leads to $c^2 = v_F^2/2$ for 2D and $c^2 = v_F^2/3$ for 3D bands.
Similarly, the analytical value for the strong coupling sound velocity follows 
from Eq.~(\ref{strongcouplingc}) 
with $\kappa_1 = \kappa_2$ and $c_{1,s} = c_{2,s}$, which leads to
$c^2=|\Delta|^2/4m|\mu|$ for 2D and $c^2=|\Delta|^2/8m|\mu|$ for 3D bands. 
Thus, we recover the Goldstone mode in both BCS and BEC limits as in the
case of the presence of only one band~\cite{jan,ohashi}. 
The numerical values (solid circles in Figs. 3a and 5a) for the sound velocity 
as a function of the dimensionless coupling $V_r$ are calculated 
from Eq.~(\ref{degeneratec}). Notice the very good agreement with the
analytical results in weak and strong coupling (dotted lines in the same figures). 
As a further consistency check, notice that this agreement is very reasonable 
since Eq.~(\ref{degeneratec}) is identical to 
the expression for the sound velocity given in Ref.\cite{jan} 
for the one-band model, with the correspondence that our coefficient $\tilde{A}$ 
has the same expression as the coefficient $A$ defined in their paper.

Notice in Fig.~3a that the sound velocity $c$ is essentially a constant for all
couplings $V_r$ in the $k_0/{k_F}_{max} \to \infty$ 
($k_0/{k_F}_{max} = 10^4$) limit. 
For smaller values of $k_0/{k_F}_{max}$ the sound velocity decreases 
as a function of $V_r$ (not shown in figure). 
We will not discuss here the dependence of $c$ on the ratio $k_0/{k_F}_{max}$
since we are mostly concerned with checking the consistency of our calculations
with the analytically tractable limits. In this case, the sound velocity
$c$ is not a good indicator that the BCS-BEC crossover is occuring in 2D. 
However, in the 3D case (see Fig. 5a), the speed of sound changes 
very fast in the neighboorhood of $\mu = 0$, thus manifesting itself as an indicator
of the crossover regime. Therefore, measurements of Goldstone mode frequency can
offer an indication of the BCS-BEC crossover possibly only in 3D 
two-band superfluids.

With the confidence of recovering the sound mode results 
for a one-band model from a two-band model with
identical bands, we proceed with the discussion of the finite frequency mode,
which is shown in Figs. 3b and 5b for the 2D and 3D cases, respectively.

The analytical value of $w_0^2$ for the weak coupling finite frequency mode
follows from Eq.~(\ref{eqn:weakcouplingw0}) for the non-degenerate case
with $V_{11} = V_{22} = V$, $N_1 = N_2 = N$, 
and $|\Delta_1| = |\Delta_2| = |\Delta|$, 
which leads to $w_0^2 = 8g_{12}|\Delta|^2/N$ for both 2D and 3D bands. 
Similarly, the analytical value for strong coupling 
follows from Eq.~(\ref{strongcouplingw0}), which leads to 
$w_0^2 = 8g_{12}|\Delta|^2/N$ for 2D, and 
$w_0^2 = 8g_{12}|\Delta|^2\sqrt{\varepsilon_F/|\mu|}/\pi N$
for 3D bands, respectively. Numerical values (solid circles in Figs. 3b and 5b) 
for the finite frequency $w_0$
as a function of the dimensionless coupling $V_r$ are calculated 
from Eq.~(\ref{degeneratew0}). Notice the very good 
agreement between the numerical results and their analytical 
counterparts in both weak and 
strong coupling limits (dotted lines in the same figures).
It is important to notice that the scales
for weak and strong coupling used in the finite
frequency plots are not the same. We scaled $w_0$ by 
$\min\{ 2 E (\mathbf{k}) \}$, which corresponds to the two-quasiparticle
excitation threshold, thus $w_0^2$ is scaled by $4|\Delta|^2$ 
since $\min\{ 2 E (\mathbf{k}) \} = 2 |\Delta|$ for $\mu > 0$,
$w_0^2$ is scaled by $4(|\mu|^2 + |\Delta|^2)$ since 
$\min\{ 2 E (\mathbf{k}) \} = 2 \sqrt{ |\mu|^2 + |\Delta|^2 }$ 
for $\mu < 0$. This choice is natural, because it indicates that
the finite frequency $w_0$ always lies below the two-quasiparticle
excitation threshold for the parameters used, meaning that the 
collective is undamped.

Notice in Figs.~3b and 5b ($k_0/{k_F}_{max} \to \infty$, 
$k_0/{k_F}_{max} = 10^4$) that the finite frequency $w_0$ changes qualitatively 
near the coupling $V_r$ where $\mu$ changes sign for both 2D and 3D cases.
We will not discuss here the dependence of $w_0$ on the ratio $k_0/{k_F}_{max}$
since we are mostly concerned with checking the consistency of our calculations
with the analytically tractable limits. However, it is important to emphasize that
the finite frequency mode is a good indicator that the BCS-BEC crossover is occuring 
in both 2D and 3D. Thus, measurements of the finite frequency $w_0$ can 
reveal BCS-BEC crossover behavior in two-band superfluids.

\section{Conclusions}
\label{sec:conclusions}

We studied the evolution of low energy collective excitations 
from weak (BCS) to strong (BEC) coupling limits
in two-band $s$-wave superfluids at $T=0$ for all intra-band coupling
strengths with ranges satisfying the diluteness condition.
We assumed that the two bands were coupled via an inter-band Josephson 
interaction. We focused on the phase-phase collective modes and showed that
there can be two undamped phase-phase modes in the evolution from weak
to strong coupling. In the weak coupling limit, we recovered 
Leggett's results corresponding to an \textit{in-phase} mode (Goldstone) mode and
an \textit{out-of-phase} mode (Leggett's mode) in the appropriate asymptotic limits.
Furthermore, we generalized Leggett's weak coupling results 
to include the BCS-BEC crossover and the strong coupling regime. 
In addition, we presented analytical results in the strong coupling limit,
in the asymptotic limit of small $g_{12}$ corresponding to weak Josephson 
coupling between the bands. All the analytical results were presented
for the cases of two-dimensional and three-dimensional bands, and
the for cases of non-degenerate and degenerate bands.

On the numerical side, we analysed fully the limit of degenerate bands, 
from which the one-band results~\cite{jan} 
can be easily recovered in the BCS, BEC, and crossover regimes 
for a 3D $s$-wave superfluid.  
The limit of degenerate bands, although less likely to be 
found in nature, provides us a good
basis for the more challenging numerical 
work for the non-degenerate case, which will be
performed in the future, as they require the self-consistent solutions of three 
simultaneous non-linear integral equations in order to determine the order 
parameters $|\Delta_1|$, $|\Delta_2|$, and the chemical potential $\mu$.

We have also briefly described in this manuscript the procedure to compute 
the amplitude-amplitude collective modes, although 
we have not discussed in detail their explicit form, 
we would like to mention that amplitude-amplitude collective modes 
are higher energy modes in comparison to lowest energy (phase-phase) collective modes
discussed here. The phase-phase and amplitude-amplitude collective 
modes are potentially important to the analysis of 
multi-component ultracold neutral Fermi gases, where
collective mode frequencies can be measured spectroscopically~\cite{litium3,litium4}. 

The results for collective modes described here (neutral) is
not strictly valid in standard (charged) condensed matter systems. 
In particular, in the BCS limit, the effect of the
Coulomb interaction is to plasmonize the 
the Goldstone mode. Furthermore, the Coulomb interaction 
modifies the velocity of the Leggett mode, but does change 
its finite frequency offset~\cite{leggett,sharapov}.
There is some experimental evidence that the Leggett mode in the BCS limit has 
been observed~\cite{ponomanev} in $\rm {MgB_2}$, but there is presently no
experimental charged two-band condensed matter system 
where the BEC limit can be reached via the tunning of an experimental parameter.
The interesting extension to the BEC limit and the crossover region
in the charged case is currently underway and will be published elsewhere. 

To conclude, the main contribution of our manuscript is to study the evolution
of the Goldstone (\textit{in-phase})  and the finite frequency 
(\textit{out-of-phase}) collective modes 
from weak (BCS) to strong (BEC) couplings in neutral two-band superfluids.
Our results are potentially relevant to multi-component ultra-cold
Fermi atoms, and can be used, in principle, to test the validity of the BCS-BEC
evolution based on extensions~\cite{carlos,perali} of 
Eagles'~\cite{eagles}, Leggett's~\cite{leggett2}, 
and Nozieres and Schmitt-Rink's~\cite{nsr} suggestions.

\section{Acknowledgements} 
We would like to thank NSF (DMR-0304380) for financial support.

\appendix
\section{Matrix Elements}

In the evaluation of the elements of the fluctuation matrix $\mathbf{M} (q)$ 
appearing in section~\ref{sec:gaussian-fluctuations}, and defined
in Eqs.~(\ref{eqn:theta-def}), we need to 
calculate the functions
$\Theta_{n,11}^{\rm qp-qh}$, $\Theta_{n,11}^{\rm qp-qp}$,
$\Theta_{n,12}^{\rm qp-qh}$, and $\Theta_{n,12}^{\rm qp-qp}$ given by
\begin{eqnarray*}
\Theta_{n,11}^{\rm qp-qh} &=& 
\sum_{\mathbf{k}} X 
\left[ \frac{|u|^2|v'|^2}{iv+E-E'} - \frac{|v|^2|u'|^2}{iv-E+E'} \right],  \\ 
\Theta_{n,11}^{\rm qp-qp} &=& 
\sum_{\mathbf{k}} Y
 \left[ \frac{|v|^2|v'|^2}{iv-E-E'} - \frac{|u|^2|u'|^2}{iv+E+E'} \right],  \\ 
\Theta_{n,12}^{\rm qp-qh} &=& 
\sum_{\mathbf{k}} X 
\left[ \frac{u^*vu'^*v'}{iv+E-E'} - \frac{u^*vu'^*v'}{iv-E+E'} \right],  \\ 
\Theta_{n,12}^{\rm qp-qp} &=& 
\sum_{\mathbf{k}} Y 
\left[ \frac{u^*vu'^*v'}{iv+E+E'} - \frac{u^*vu'^*v'}{iv-E-E'} \right].
\end{eqnarray*}
In the previous expressions the indices ${\rm qp-qh}$ and ${\rm qp-qp}$ classify
quasiparticle-quasihole and quasiparticle-quasiparticle terms, 
respectively. Furthermore, we used the following simplified notation:  
the kinetic energies $\xi_n(\mathbf{k}) \to \xi$,
$\xi_n(\mathbf{k}+\mathbf{q}) \to \xi'$;  
the quasiparticle energies $E_n(\mathbf{k}) \to E$,
$E_n(\mathbf{k}+\mathbf{q}) \to E'$; 
the order parameters $\Delta_n(\mathbf{k}) \to \Delta$,
$\Delta_n(\mathbf{k}+\mathbf{q}) \to \Delta'$;
the symmetry factors 
$\Gamma_n(\mathbf{k}) \to \Gamma$,
$\Gamma_n(\mathbf{k}+\mathbf{q}/2) \to \Gamma'$;
and the Fermi functions
$f_n(E_n(\mathbf{k})) \to f$,
$f_n(E_n(\mathbf{k}+\mathbf{q})) \to f'$. 
We also made use of the definition of the first 
coherence factor 
\begin{equation}
|u|^2 = \frac{1}{2}(1 + \frac{\xi}{E}), 
\end{equation}
[$|u'|^2=(1+\xi'/E')/2$],
the second coherence factor
\begin{equation}
|v|^2 = \frac{1}{2}(1-\frac{\xi}{E})
\end{equation}
[$|v'|^2=(1-\xi'/E')/2$],
and the phase relation between them
\begin{equation}
u^* v = \frac{\Delta}{2E}
\end{equation}
[$u'^* v' = \Delta'/2E'$].
Finally, we used the notation
$X =(f-f')\Gamma'^2$ and $Y = (1-f-f')\Gamma'^2$
to indicate the combinations of Fermi functions $(f, f')$ 
and symmetry coefficients $(\Gamma, \Gamma')$ appearing in
the quasiparticle-quasihole, and quasiparticle-quasiparticle terms,
respectively.

\section{Expansion Coefficients}

From the rotated fluctuation matrix $\widetilde \mathbf{M}$
expressed in the amplitude-phase basis as defined in 
section~\ref{sec:collective-modes}, we can obtain the expansion
coefficients necessary to calculate the collective modes
at $T = 0$. In the long-wavelength $(\mathbf{q} \to \mathbf{0})$, and low
frequency limit $(w \to 0)$ the matrix $\widetilde \mathbf{M}$
defined in Eq.~(\ref{eqn:fluctuation-matrix}) is fully determined 
by the knowledge of the matrix elements $x_n$, $y_n$ and $z_n$.
Note that, this expansion requires 
\begin{equation}
w,|\mathbf{q}|^2/m_n \ll \min\{E_1(\mathbf{k}), E_2(\mathbf{k})\}
\end{equation}
and these coefficients are valid for all couplings for $s$-wave pairing.

In all the expressions below we use the following simplifying notation 
$\dot{\Gamma}=\partial\Gamma/\partial k$,
$\dot{\xi}=\partial\xi/\partial k$,
$\ddot{\Gamma}=\partial^2\Gamma/\partial k^2$, 
$\ddot{\xi}=\partial^2\xi/\partial k^2$, and $k=\vert \mathbf{k} \vert$.

The coefficients necessary to obtain the matrix element $x_n$ are
\begin{equation}
A_n  = -g_{nn} - \sum_{\mathbf{k}}\frac{\xi^2}{2E^3}\Gamma^2 , 
\end{equation}
corresponding to the $(\mathbf{q} = 0, w = 0)$ term,
\begin{eqnarray}
C_n & = & \sum_{\mathbf{k}}\frac{1}{8E^5}
{\Big\lbrace}
\ddot{\xi}\xi(E^2-3\Delta^2)\Gamma^2 + \xi^2(2\Delta^2-\xi^2) \Gamma\ddot{\Gamma}
\nonumber \\
&-&\dot{\xi}^2\left[ E^2-10\Delta^2\left( 1-\frac{\Delta^2}{E^2} \right) 
\right]\Gamma^2 \cos^2\alpha  \nonumber \\
&+&2\dot{\xi}\xi\left[ E^2+\Delta^2\left( 1-10\frac{\xi^2}{E^2} \right) \right] 
\Gamma\dot{\Gamma}\cos^2\alpha \\
&-&\left[ \xi^2 (\xi^2-7\Delta^2) - 5\Delta^4\left( 1-2\frac{\xi^2}{E^2} \right) \right] 
\dot{\Gamma}^2\cos^2\alpha \nonumber
{\Big\rbrace}, 
\end{eqnarray}
corresponding to the $|\mathbf{q}|^2$ term with $\alpha$ being the angle
between $\mathbf{k}$ and $\mathbf{q}$;
\begin{equation}
D_n = \sum_{\mathbf{k}}\frac{\xi^2}{8E^5}\Gamma^2,
\end{equation}
corresponding to the $w^2$ term,
and
\begin{equation}
F_n = -\sum_{\mathbf{k}}\frac{\xi^2}{32E^7}\Gamma^2,
\end{equation}
corresponding to the $w^4$ term.

The coefficients necessary to obtain the matrix element $y_n$ are
\begin{equation}
P_n = -g_{nn} - \sum_{\mathbf{k}}\frac{1}{2E}\Gamma^2,
\end{equation}
corresponding to the $(\mathbf{q} = 0, w = 0)$ term,
\begin{eqnarray}
Q_n =\sum_{\mathbf{k}}\frac{1}{8E^5} 
\Big\lbrace
\ddot{\xi}\xi E^2\Gamma^2
-\dot{\xi}^2( E^2-3\Delta^2 )\Gamma^2 
- \xi^2E^2 \Gamma\ddot{\Gamma} \nonumber \\
+ \left[ 
\dot{\xi}\xi( 2E^2-6\Delta^2 ) \Gamma\dot{\Gamma}   
 - \xi^2( \xi^2 - 2\Delta^2) \dot{\Gamma}^2 
\right]
\cos^2 \alpha  
\Big\rbrace, 
\end{eqnarray}
corresponding to the $|\mathbf{q}|^2$ term with $\alpha$ being the angle
between $\mathbf{k}$ and $\mathbf{q}$;
\begin{equation}
R_n = \sum_{\mathbf{k}}\frac{1}{8E^3}\Gamma^2, \\
\end{equation}
corresponding to the $w^2$ term,
and 
\begin{equation}
T_n = -\sum_{\mathbf{k}}\frac{1}{32E^5}\Gamma^2, \\
\end{equation}
corresponding to the $w^4$ term.

The coefficient necessary to obtain the matrix element $z_n$
is
\begin{equation}
B_n = \sum_{\mathbf{k}}\frac{\xi}{4E^3}\Gamma^2, 
\end{equation}
corresponding to the $w$ term,
and
\begin{equation}
H_n = \sum_{\mathbf{k}}\frac{\xi}{16E^5}\Gamma^2, 
\end{equation}
corresponding to the $w^3$ term.

\end{document}